\documentclass[a4paper,11pt,fleqn]{article}
\pdfoutput=1
\usepackage{jheppub}

\usepackage{graphicx}
\usepackage[figuresright]{rotating}
\usepackage{bm,amsmath,amssymb}
\usepackage[mathscr]{eucal}
\usepackage{color}
\usepackage{multirow}
\usepackage{subcaption}

\long\def\comment#1{ }
\newcommand{\eqn}[1]{Eq.~\eqref{#1}}
\newcommand{\beq}{\begin{equation}}
\newcommand{\eeq}{\end{equation}}

\newcommand{\dif}{{\rm{d }}}
\newcommand{\rmd}{{\rm{d }}}

\newcommand{\ombr}{\omega_{\textrm{br}}}
\newcommand{\tbr}{t_{\textrm{br}}}
\newcommand{\omc}{\omega_c}

\newcommand{\abar}{\bar{\alpha}_s}
\newcommand{\amed}{{\alpha}_{s,\text{med}}}
\newcommand{\baramed}{\bar{\alpha}_{s,\text{med}}}

\newcommand{\zcut}{z_{\textrm{cut}}}
\def\om{\omega}
\def\qhat{\hat{q}}
\def\th{\theta}

\usepackage{color}
\definecolor{darkgreen}{rgb}{0,0.5,0}
\definecolor{darkblue}{rgb}{0,0,0.7}
\definecolor{darkred}{rgb}{0.5,0,0.0}
\definecolor{darkorange}{rgb}{0.8,0.4,0.0}

\makeatletter
\g@addto@macro\bfseries{\boldmath}
\makeatother

\title{Jet radiation in a longitudinally expanding medium}

\author{P.~Caucal,}
\author{E.~Iancu,}
\author{and G.~Soyez}
\affiliation{Institut de Physique Th\'{e}orique, Universit\'{e} Paris-Saclay, CNRS, CEA, F-91191, Gif-sur-Yvette, France}

\emailAdd{paul.caucal@ipht.fr}
\emailAdd{edmond.iancu@ipht.fr}
\emailAdd{gregory.soyez@ipht.fr}
\abstract{
In a series of previous papers, we have presented a new approach,
based on perturbative QCD, for the evolution of a jet in a dense
quark-gluon plasma.
In the original formulation, the plasma was assumed to be homogeneous
and static.
In this work, we extend our description and its Monte Carlo
implementation to a plasma obeying Bjorken longitudinal expansion.
Our key observation is that the factorisation between vacuum-like and
medium-induced emissions, derived in the static case, still holds for
an expanding medium, albeit with modified rates for medium-induced
emissions and transverse momentum broadening, and with a modified
phase-space for vacuum-like emissions.
We highlight a scaling relation valid for the energy spectrum of
medium-induced emissions, through which the case of an expanding
medium is mapped onto an effective static medium.
We find that scaling violations due to vacuum-like emissions and
transverse momentum broadening are numerically small.
Our new predictions for the nuclear modification factor for jets
$R_{AA}$, the in-medium fragmentation functions, and substructure
distributions are very similar to our previous estimates for a static
medium, maintaining the overall good qualitative agreement with
existing LHC measurements.
In the case of $R_{AA}$, we find that the agreement with the data is
significantly improved at large transverse momenta
$p_T\gtrsim 500$~GeV after including the effects of the nuclear parton
distribution functions.}

\begin{document}
\maketitle

\section{Introduction}

The physics of jet quenching --- a generic denomination for the
modifications that a jet produced in the dense 
environment of an ultrarelativistic heavy-ion collision undergoes when
propagating through and interacting with this surrounding medium --- represents one of the main
tools at our disposal for experimentally probing the quark-gluon
plasma, the state of QCD matter at high partonic densities. There is
by now overwhelming evidence, notably from Au+Au collisions at RHIC
and Pb+Pb collisions at the LHC, for strong nuclear modifications of
the jet properties. In order to draw the appropriate lessons from
these data,
the experimental efforts must be accompanied by theoretical progress, aiming at understanding the 
jet-medium interactions from first principles.

Perturbative QCD offers a suitable framework for systematic
first-principles studies.  Its applicability to the problem of jet
quenching is by no means obvious: the relevant phenomena involve
widely separated scales, including relatively soft ones, like the
temperature of the medium, which is not much higher than
the QCD confinement scale. It is not straightforward either: even 
when the coupling is weak, the high-parton densities
entail collective phenomena, like multiple scattering and multiple
medium-induced emissions, which call for all-order resummations. Last
but not least, irrespective of the value of the coupling, there are
aspects of the dynamics --- like the geometry of the collisions, the
rapid longitudinal expansion of the partonic medium created in the
wake of the collision, or the hadronisation of the jet constituents at
late times --- which are intrinsically non-perturbative.
In order to encompass such complex phenomena, 
a perturbative setup should be supplemented by an
(as-small-as-possible) amount of non-perturbative modelling.

In this paper, we work in the context of the perturbative picture
proposed and developed in
Refs.~\cite{Caucal:2018dla,Caucal:2019uvr,Caucal:2020xad} (see
also~\cite{Caucal:2020isf} for an extensive review and additional
calculations), which itself builds upon a series of earlier
first-principle developments.
The physical picture underlying our approach is anchored in a remarkable property
  emerging from perturbative QCD: the parton cascades
are factorised between {\it vacuum-like emissions (VLEs)}, which are triggered by the virtuality of the initial 
parton and occur at early times, and  {\it medium-induced emissions (MIEs)}, which are triggered by 
collisions with the plasma constituents and can occur anywhere inside the medium. 

Originally justified in a double-logarithmic approximation, in which successive VLEs are strongly 
ordered in both energies and emission angles~\cite{Caucal:2018dla}, this factorised picture has 
later been argued to remain valid in a less restrictive, single logarithmic, approximation, 
which assumes angular ordering alone~\cite{Caucal:2019uvr}.
The treatment of the MIEs is based on the BDMPSZ 
approach~\cite{Baier:1996kr,Baier:1996sk,Zakharov:1996fv,Zakharov:1997uu,Baier:1998kq,Baier:1998yf,Zakharov:1998wq,Wiedemann:1999fq,Wiedemann:2000za}, 
which takes into account the coherence effects associated with 
multiple scattering during the quantum formation of an emission. Multiple branching, which 
becomes important for relatively soft MIEs, has been included by iterating the BDMPSZ rate, 
as proposed in Refs.~\cite{Blaizot:2012fh,Blaizot:2013hx,Blaizot:2013vha} (see also~\cite{Baier:2000sb,Arnold:2002zm,Arnold:2002ja,Jeon:2003gi,Kurkela:2011ti,Apolinario:2014csa,Kurkela:2014tla,
Arnold:2015qya,Arnold:2016kek} for related work).

Under these assumptions, both the vacuum-like emissions and the
medium-induced ones are described by Markovian branching processes,
albeit with different ordering variables: the emission angle for the
VLEs and, respectively, the physical time for the MIEs. This probabilistic
description allowed us to develop a Monte-Carlo (MC) event generator
which includes both types of branching processes in a simple modular
structure which translates the factorisation between VLEs and
MIEs. Its first applications to the phenomenology of jet quenching
turned out to be encouraging~\cite{Caucal:2019uvr,Caucal:2020xad},
despite the rather crude description of the medium itself.

Indeed, in all these applications, the medium was assumed to be a homogeneous and stationary 
slice of plasma of longitudinal width  $L$, the distance travelled by
the jet  inside the medium. Furthermore, the interactions between the
jet and the plasma have so far been fully characterised by a constant 
 ``jet quenching parameter'' $\hat q$, the rate for transverse momentum broadening via collisions.
This oversimplified picture neglects important phenomena like the longitudinal and radial
expansions of the medium, the medium geometry and the associated distribution of the hard interaction point (hence, of the effective medium size $L$), and the medium response to the jet (e.g.\ the soft hadrons
from the underlying event which are dragged by the jet and cannot be distinguished from
genuine jet components). In this paper we shall improve our approach
by taking into account the longitudinal expansion of the medium. 
 
More concretely, our objective is twofold. On the conceptual side, we will demonstrate that
the factorised picture for in-medium jet radiation that we put forward in~\cite{Caucal:2018dla,Caucal:2019uvr,Caucal:2020xad,Caucal:2020isf} remains valid for a medium undergoing 
Bjorken longitudinal expansion~\cite{Bjorken:1982qr}, modulo a few
suitable adaptations to account for the expanding medium.
On the phenomenology side, we will check via explicit
Monte Carlo simulations that the qualitative agreement that we previously observed
between our predictions and the experimental data for a few selected observables
remains equally good after including the effects of the longitudinal expansion.

The first modification in our factorised picture is quite obvious: the rate for transverse
momentum broadening is now controlled by a time-dependent transport coefficient,
which for the Bjorken (isentropic) expansion takes the form of a power law:\footnote{A 
more general behaviour proportional to $(t_0/t)^\gamma$, with $\gamma \le 1$, will be considered in our theoretical
arguments, but the most common choice $\gamma=1$ will be privileged
in the applications to phenomenology.}
$\hat q (t)=\hat q_0 ({t_0}/{t})$. The initial value $\hat q_0$ and the initial time $t_0$
can be treated as free parameters, or, alternatively, they can be both related to the
gluon saturation momentum $Q_s$ in the incoming nuclei~\cite{Baier:2000sb}. In this paper,
we shall adopt the second viewpoint, which has the fringe benefit of  preserving the same number
of free parameters as for a static medium ($\hat q$ gets replaced by $Q_s$).  
In practice, the ratio $L/t_0$ is quite large --- $L/t_0\gtrsim 20$
for the typical values we will consider  ---, showing that the dilution 
of the medium via longitudinal expansion is a sizeable effect, with potentially large consequences.

The other modifications associated with the expansion of the medium are less obvious,
as they refer to radiation processes, which are non-local in time. Since the VLEs occur, by definition,
``like in the vacuum'', one may think that they are insensitive to the  properties of the medium,
 in particular to its expansion. As already explained in~\cite{Caucal:2018dla}, this
is  not true: the phase-space for the VLEs occurring inside the plasma is reduced by
medium rescattering. This reduction avoids the overlap in phase-space between VLEs and
MIEs and thus lies at the heart of our argument in favour of factorisation. 
We show that the argument underlying this 
particular constraint (the boundary of the ``vetoed region'') can be transposed from the
static medium to the longitudinally expanding one --- so in particular, the factorisation
property remains true.

Concerning the MIEs, we adopt the same strategy as in the case of a static medium, namely
we provide a faithful description only for the relatively soft emissions, with energies 
$ \omega\ll \omega_c\simeq\hat q(L) L^2$. Such emissions are the most interesting
for our purposes, as they control important phenomena like the energy lost by the jet, or 
the nuclear effects on the jet fragmentation and several substructure observables.
A key feature of the soft MIEs is that they have short formation times, $t_f \ll L$, 
meaning that  multiple emissions are important and must be resummed to all orders.
The fact that $t_f \ll L$ greatly simplifies this
resummation~\cite{Blaizot:2012fh,Blaizot:2013hx,Blaizot:2013vha} since
it implies that the soft emissions 
can be effectively  treated as  instantaneous and resummed by iterating
an emission rate --- in turn related to the low-energy approximation of the BDMPSZ spectrum for
an expanding medium~\cite{Baier:1998yf,Zakharov:1998wq}.
By the same argument,  the emission rate for an expanding medium 
is found to be the same as for a static medium, with $\hat{q}$ replaced
  by an instantaneous $\hat{q}(t)$ measured at the emission time.

This last observation implies that the medium-induced branching processes for the
expanding and the static media,  respectively,  can be exactly mapped onto each other via 
a redefinition of the time variable.
This mapping entails a very useful scaling relation
between the parton distributions generated via MIEs in the two cases:
their longitudinal distributions
become identical if the jet quenching parameter for the static medium is
chosen to obey $\hat q_\text{stat}\simeq 4\hat q(L)$ (cf.\
  Eq.~(\ref{qeff}) for the exact definition). We note that this scaling differs from the
one  originally observed\footnote{With our present conventions,
the scaling in Ref.~\cite{Baier:1998yf} amounts to $\hat q_\text{stat}=2\hat q(L)$; see also 
\eqn{aveq} below.} in Ref.~\cite{Baier:1998yf} and 
studied numerically in Refs.~\cite{Salgado:2002cd,Salgado:2003gb}:
that property was derived for the average energy loss by the leading
parton, a quantity controlled by gluon emissions with relatively large
energies $ \omega\sim \omega_c$, whereas our scaling is strictly valid
for $ \omega\ll\omega_c$. %
It was recently observed~\cite{Adhya:2019qse} that the full BDMPSZ spectrum for
an expanding medium~\cite{Baier:1998yf,Zakharov:1998wq} actually admits two
scaling properties: one which becomes exact at low energies $ \omega\ll \omega_c$,
and one which is better verified at larger energies $ \omega\sim \omega_c$.
Since our approach uses an approximate version of the 
BDMPSZ spectrum valid for soft emissions, it only reproduces the
low-energy scaling, which is exactly  satisfied in our soft limit.

However, this scaling is violated by the other ingredients of the dynamics ---
the transverse momentum broadening and the VLEs ---, because of their different functional
dependencies upon $\hat q(t)$. These violations are particularly interesting, since they are
consequences of the longitudinal expansion which cannot be ``scaled out'', i.e. cannot be
identically reproduced by a well chosen static medium, with $\hat
q_\text{stat}\simeq 4\hat q(L)$.
In this paper we shall study these scaling violations via a combination
of numerical (Monte Carlo) and analytic methods. 
By comparing results for the expanding medium and for the ``equivalent'' 
static one with $\hat q_\text{stat}\simeq 4\hat q(L)$, we will identify two types of
scaling violations: a slight reduction in the phase-space for the VLEs, and a (similarly
small) reduction in the transverse momentum broadening of the soft gluons.
Physically, they reflect the fact that the collisions are less effective in an 
expanding medium, which is rapidly diluting, than in a static one, even if the latter is ``well-tuned''.

The main conclusion emerging from our analysis is that the scaling
violations only have a small effect, at the level of a few percent, on
all the quantities that we have investigated.  These include the
average energy loss by the jet (cf.\ Sect.~\ref{sec:scaling}) and the
observables that we had previously computed for the static
case~\cite{Caucal:2019uvr,Caucal:2020xad} and that we shall here
recompute for the longitudinally-expanding plasma (see
Sect.~\ref{sec:pheno}): the nuclear modification factors (a.k.a.\
$R_{AA}$ ratios) for the inclusive jet production, the jet
fragmentation function, and the Soft
Drop~\cite{Dasgupta:2013ihk,Larkoski:2014wba} distributions for
$z_g$~\cite{Larkoski:2015lea} and $\th_g$.

For all these observables, the nuclear effects are dominated by the
energy lost by the jet via soft MIEs at large angles. %
The (almost-exact) scaling property that emerges from our analysis
guarantees that an equally-good description of these observables can
be obtained for an expanding medium and for a static one.
Indeed, our Monte Carlo predictions for an expanding medium are as
good as those obtained in our previous
papers~\cite{Caucal:2019uvr,Caucal:2020xad} for a static medium.
We find a qualitative, and even semi-quantitative, agreement between
our predictions and the respective LHC data for all
the observables that we considered and for several (physically-reasonable)
values for our only three free parameters.

In the case of the $R_{AA}$ ratio for inclusive jet production, we find that the agreement with
the LHC data is considerably improved at large transverse momenta
$p_T\gtrsim 500$~GeV after including the effect of 
nuclear parton distribution functions~\cite{Eskola:2016oht} for the 
hard process which initiates the jets.
Physically, this yields a suppression of the quark distribution in the
incoming nuclei compared to a proton at large values $x\sim 1$ for the
longitudinal momentum fraction~\cite{Malace:2014uea}.

The structure of this paper is as follows. In Sect.~\ref{sec:ps}, after briefly recalling the underlying 
physics assumptions and the general structure of our theoretical description, we discuss the modifications 
which occur in the formalism after taking into account the longitudinal expansion of the medium. 
In Sect.~\ref{sec:mc}, we describe the practical consequences of these modifications for 
the Monte-Carlo implementation of our approach.
In Sect.~\ref{sec:scaling}, we discuss the scaling property which relates parton distributions created
via medium-induced emissions in a longitudinally-expanding medium to
that in an ``equivalent'' static medium. After explaining the theoretical basis of this scaling and of its violations by the
full dynamics, we present numerical tests together with analytic calculations which illustrate
the scaling quality. In Sect.~\ref{sec:pheno},
we present MC calculations for the observables aforementioned. At several places, we compare
the respective predictions for an expanding medium and the ``equivalent'' static medium, in order to
emphasise their strong similarity. We present for the first time
in our picture results including the nuclear PDFs
and predictions for the distribution of the Soft Drop subjet separation angle $\th_g$.
We present our conclusions in Sect.~\ref{sec:conc}.

\section{Parton showers in a longitudinally-expanding medium}
\label{sec:ps}

In this section we shall describe the generalisation of our physical picture in 
Refs.~\cite{Caucal:2018dla,Caucal:2019uvr,Caucal:2020xad} to the case of a plasma which
obeys longitudinal expansion. The consequences of this expansion for the
transverse momentum broadening and for the medium-induced radiation have been
explored at length in the literature (see e.g. Refs.~\cite{Baier:1998yf,Zakharov:1998wq,Arnold:2008iy,Salgado:2002cd,Salgado:2003gb,Adhya:2019qse,Iancu:2018trm}
for approaches similar to ours). In what follows we shall build upon such previous studies to
incorporate the medium expansion in our unified description for the in-medium parton showers,
including both vacuum-like and medium-induced emissions.

\subsection{The physical picture for a static plasma}\label{sec:recall-static}

Let us first recall the main approximations underlying our effective
theory for jet radiation, as originally developed for a static medium
in Refs.~\cite{Caucal:2018dla,Caucal:2019uvr,Caucal:2020xad}. The main
feature of our picture is a factorisation in time between the {\it
  vacuum-like emissions} (VLEs), triggered by the initial virtuality
of the leading parton, and the {\it medium-induced emissions} (MIEs)
triggered by collisions in the plasma. It is convenient to describe
separately each stage in the temporal development of the parton
cascades.

\texttt{(i)} {\it VLEs inside the medium.}
The VLEs occurring inside the medium are computed in a leading
logarithmic approximation which implies, in particular, strong
ordering in the emission angles. Whereas this ordering is natural for
jets propagating in the vacuum, due to the (quantum and colour)
coherence of the partonic sources, it is non-trivial for jets
propagating through a dense medium, where colour coherence can be
washed out by the collisions with the medium constituents. Yet, we
have shown in~\cite{Caucal:2018dla} that angular ordering is preserved
for the VLEs developing inside the medium, mainly due to their
sufficiently short formation time. The only effect of the medium
during this stage is a kinematical constraint which reduces the
phase-space available for radiation (cf.\ the region labelled ``inside medium''
in Fig.~\ref{Fig:phase-space} below).

\texttt{(ii)} {\it MIEs through the medium.}
After formation, the partons produced via VLEs propagate through the
medium and act as sources for the next stage, medium-induced
radiation.
In practice, the MIEs are iterated by following a Markovian
branching process, with the branching rate tuned to match the BDMPSZ
spectrum for a single gluon emission with relatively small energy
$\omega\ll \omc\equiv \hat q L^2/2$.
This probabilistic picture provides a faithful description of the soft gluon emissions
with energies $\omega \lesssim \ombr\equiv \abar^2\omc$, for which multiple branching
is expected to be important. Such soft gluons propagate at large angles 
(typically, outside the jet boundary) and are the main source of jet energy
loss.
The somewhat harder emissions, with intermediate energies
$\ombr < \omega < \omc$,
generally remain
inside the jet, affecting its substructure, and act as additional
sources for softer radiation (hence for energy loss).
Such emissions are rare, so they are correctly described by the single emission limit of
our branching process, at least so long as $\omega\ll \omc$. Finally, the very hard emissions 
with $\omega\gtrsim \omc$ are strongly suppressed, since they require a relatively hard scattering; 
we have modelled this suppression by a sharp upper cutoff at $\omega=
\omc$ on the medium-induced spectrum.

\texttt{(iii)}  {\it VLEs outside the medium.}
The partons produced inside the medium, via either VLEs or MIEs, are still virtual, with
a minimal virtuality (or transverse momentum) of order $k_\perp^2\sim \sqrt{\hat q\omega}$, as 
introduced by collisions during the formation time. They will evacuate this virtuality (down to the
hadronisation scale $\Lambda$) via parton emissions outside the medium, which follow the standard
vacuum angular-ordered pattern, with one noticeable exception: the
very first emission outside the medium can occur at {\it any} angle.
This happens because this emission has been sourced by partons which have crossed 
the medium along a large distance, of order $L$, and which
have lost their coherence via rescattering, so they can be seen as independent.
This is conceptually important as it opens the angular phase-space
beyond what would normally happen in a vacuum parton cascade.

\subsection{Basic characterisation of a longitudinally expanding
  plasma}\label{sec:expanding-basic}

Our main purpose in this paper is to generalise the above
picture to the case of a plasma undergoing longitudinal expanding according to
the Bjorken picture~\cite{Bjorken:1982qr}, i.e.\ such that the distribution of particles is
boost-invariant. This is a reasonably good approximation for the bulk matter created in 
heavy-ion collisions in the so-called ``central plateau'' region~\cite{Bjorken:1982qr}
and, in particular, at midrapidities.
For such a plasma, the parton density $\rho$ depends only upon the
proper time $\tau\equiv\sqrt{t^2-z^2}$ (with $z$ referring to the collision
axis), and so does the jet quenching parameter $\hat q$, which is proportional
to $\rho$ (at least in perturbation theory).

In what follows, we shall restrict ourselves to jets propagating at
central rapidities:\footnote{The generalisation to more general rapidities (still within the central plateau) could be obtained by following~\cite{Iancu:2018trm}.} $\eta\simeq 0$, or $z\simeq 0$, so we can identify $\tau\equiv t$.
As explained e.g.\ in Appendix~A of Ref.~\cite{Baier:1998yf}, the time dependence of the parton
density, hence of $\hat q$, can be easily computed for the case of an
isentropic flow; one finds
\begin{equation}\label{qt}
\hat q (t)\,\simeq\,\hat q_0 \left(\frac{t_0}{t}\right)^\gamma,
\qquad\mbox{with}\quad\gamma\equiv 3v_s^2=\frac{1}{1+\Delta_1/3}\,.
\end{equation}
Here $v_s$ denotes the sound velocity, which would be equal to
$1/\sqrt{3}$ (implying $\gamma=1$) for an ideal gas. The parameter
$\Delta_1$, which measures the deviation from the ideal gas limit, is
positive in perturbation theory and of order ${\alpha_s^2}$, implying
that $\gamma$ is close to, but smaller than, one. In practice, we treat this power as a free parameter,  for which we consider 
various values $0\le \gamma\le 1$, with $\gamma=0$ corresponding to
the static case. Note also that $\hat q_0 $ (and hence $\hat q (t)$) is
proportional to the Casimir $C_R$ for the colour representation of the
parton; in what follows we reserve the simple notation $\hat q$
for the case where the parton is a gluon. The corresponding quantity
for a quark is $\hat q_F=(C_F/C_A)\hat q$.

The initial time $t_0$ in \eqn{qt} is, roughly speaking, the time
after which a partonic medium has been created by the collision. On
physical grounds, this is expected to be the time after which the
small-$x$ gluons from the wavefunctions of the incoming nuclei have
been liberated by the collision (see e.g.\ the discussion 
in~\cite{Baier:2000sb}).  These gluons have transverse momenta of the
order of the nuclear saturation momentum $Q_s^2\equiv Q_s^2(A,x)$ and
longitudinal wavelengths $\lambda_z = 1/p_z\lesssim 1/Q_s$. Hence one
can estimate $t_0\simeq 1/Q_s$, corresponding to the time required for
the two nuclei to cross each other along a distance given by the
maximal $\lambda_z$ value. Similarly, we
have $\hat q_0 t_0\simeq Q_s^2$, since transverse momentum broadening
starts at time $t_0$ with the intrinsic momentum of the liberated
gluons.
In what follows, we shall use these simple estimates to replace the 2
parameters $\hat q_0$ and $t_0$ by a single one, $Q_s^2$, which takes
the place of the (time-independent) jet quenching parameter $\hat q $
used for a static medium.  Some representative values that we shall
later use for applications are $Q_s=1\div 2$~GeV (for a Pb nucleus at
$x\sim 10^{-3}$), implying $t_0=1/Q_s=0.1\div 0.2$~fm and
$\hat q_0=Q_s^3=5\div 40$~GeV$^2$/fm.

\subsection{Vacuum-like-emissions (VLEs) in the longitudinally
  expanding plasma}
  \label{sec:VLEs-in-expanding}

The original hard process (generally, a $2\to 2$ partonic scattering)
giving rise to the leading parton with initial energy $E$ occurs very
fast, over a time $t\sim 1/E\ll t_0$, that can be safely set to zero
when studying the subsequent evolution of the jet and of the medium.
The leading parton is typically produced with a large, time-like,
virtuality that it evacuates via successive emissions. The early
emissions are typically hard and thus have short formation times
$t_f\simeq 2\omega/k_\perp^2 \simeq 2/(\omega\theta^2)$, with $\omega$
the energy of the emitted gluon, $k_\perp\simeq \omega\theta$ its
transverse momentum and $\theta$ its angle relative to the jet axis.
These emissions can occur either before the formation of the medium, 
$t_f < t_0$, i.e.\ truly in the vacuum, of within within
the (time-dependent) medium, when $t_f>t_0$.

In our perturbative picture, an emission is considered 
vacuum-like (as opposed to medium-induced) provided its  transverse momentum $k_\perp$
is much larger than the respective momentum that would be acquired via collisions during
formation: $k_\perp^2 \gg \langle k_\perp^2 \rangle (t_f,t_0)$, where
\begin{equation}\label{ktbroad}
 \langle k_\perp^2 \rangle (t,t_0) \equiv 
\int_{t_0}^{t}\dif t'\,\hat q (t')\,=\,\hat q_0  t_0^\gamma\,\frac{t^{1-\gamma}-t_0^{1-\gamma}}
{1-\gamma}\,=\,\hat q (t)t\,\frac{1-(t_0/t)^{1-\gamma}}
{1-\gamma}\quad \underset{\gamma \to 1}{\longrightarrow}\ \hat q_0  t_0\ln\frac{t}{t_0}\,,
\end{equation}
where we repeatedly used \eqn{qt} for $\hat q(t)$. The condition 
$k_\perp^2 \gg \langle k_\perp^2 \rangle (t_f,t_0)$ introduces a restriction on the phase-space
$(\omega,\,\theta)$
for VLEs,  conveniently formulated as a low-energy boundary $\omega >
\omega_0(\theta)$, 
where
$\omega_0(\theta)$ is implicitly defined by
\begin{equation}\label{omega0}
\omega_0^2\theta^2=\langle k_\perp^2 \rangle (t_f, t_0)\qquad\mbox{with}\qquad
t_f=\frac{2}{\omega_0\theta^2}\,.\end{equation}
This equation
can be further simplified without compromising our leading-logarithmic
accuracy.
Indeed, the phase-space for VLEs is best represented in logarithmic variables, say
$\ln\omega$ and $\ln(1/\theta)$, to properly account for the logarithmic, soft and collinear,
singularities of the bremsstrahlung spectrum. Considering $\gamma=1$
for definiteness, Eqs.~\eqref{ktbroad}--\eqref{omega0}  imply $\omega_0^2\theta^2=Q_s^2 \ln({t_f}/{t_0})$,
hence $\ln(\omega_0^2\theta^2)=\ln Q_s^2 + \ln \ln({t_f}/{t_0})$. The second term in
the r.h.s.\ is slowly varying and can safely be neglected.
Similarly, for $\gamma < 1$ and $t_f\gg t_0$,
Eqs.~\eqref{ktbroad}--\eqref{omega0}  imply  $\ln(\omega_0^2\theta^2)=\ln[\hat q (t_f)t_f]
+\ln[1/(1-\gamma)]$ and the constant shift $\ln[1/(1-\gamma)]$ can be ignored.
Hence, to the accuracy of interest, the solution $\omega_0(\theta)$ to \eqn{omega0} 
 for any $\gamma \le 1$ can be written as\footnote{In order to recover the known
result for the static medium in the limit $\gamma\to 0$, one should also remember that, in that
limit, one should replace $Q_s\to \hat q^{1/3}$; then \eqn{omega0fin} reduces to
$\omega_0^3\theta^4=2\hat q$, or $k_\perp^2 = \sqrt{2\hat q\omega}$.}
\begin{equation}\label{omega0fin}
\om\ge  \omega_0(\th)\equiv \left(\frac{2^{1-\gamma}\qhat_0t_0^{\gamma}}{\th^{4-2\gamma}}\right)^{\frac{1}{3-\gamma}}=\left(\frac{2^{1-\gamma}\qhat(L)L^\gamma}{\th^{4-2\gamma}}\right)^{\frac{1}{3-\gamma}}
= 2^{\frac{1-\gamma}{3-\gamma}}\theta^{{-}\frac{4-2\gamma}{3-\gamma}}\,Q_s,
\end{equation}
where in the last step we have also used $\hat q_0 t_0=Q_s^2$ and $t_0=1/Q_s$ to
simplify the result.

\begin{figure}[t] 
  \centering
    \includegraphics[width=0.55\textwidth]{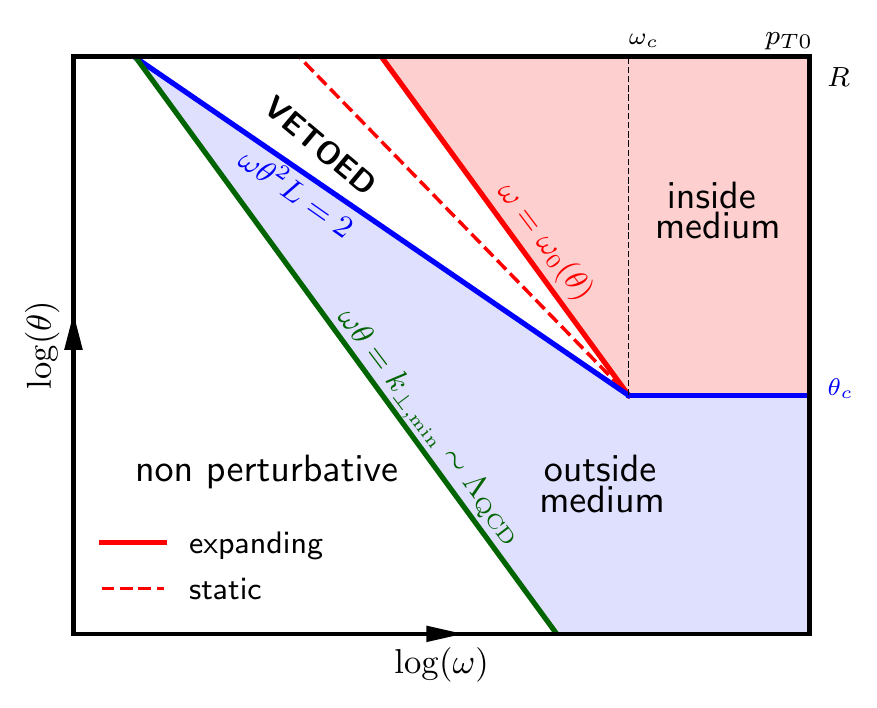}
  \caption{\small Double logarithmic phase space for VLEs in an
    expanding medium. The in-medium boundary (plain red line) is set by
    Eq.~\eqref{omega0fin} and is our reference choice for the
    Monte-Carlo implementation. The (red) dotted line represents the boundary of
    the ``equivalent'' static medium (see the discussion in
    Sect.~\ref{sec:scaling}).}
    \label{Fig:phase-space}
\end{figure}

This restriction  $\omega > \omega_0(\theta)$ applies of course only to the VLEs occurring
inside the medium, i.e.\ such that the respective formation times satisfy $t_f<L$, or $\omega>
\omega_L(\theta)\equiv 2/(L\theta^2)$. We conclude that, as in the case of a static medium,
the phase-space for VLEs inside the medium contains a vetoed region, at $\omega_L(\theta)
< \omega < \omega_0(\theta)$. This region ends at the point $(\omega_c,\,\theta_c)$
where the two boundaries intersect with each other (see Fig.~\ref{Fig:phase-space}):
\begin{equation}\label{omegac} \omega_c\,=\,\frac{ \hat{q}(L)L^2}{2}\,,\qquad
\theta_c \,=\,
  \frac{2}{\sqrt{{\hat q}(L) L^3}}\,,
  \end{equation}
which look formally similar to the respective expressions for the static medium,
except for the replacement of the time-independent jet quenching parameter $\hat q$ by its value
$\hat q (L)$ at $t=L$. This replacement has important consequences.
For example, for $\gamma=1$, \eqn{omegac}
implies $\omega_c = Q_s^2L/2$ and $\theta_c=2/(Q_sL)$, meaning in
particular that $\omega_c(L)$ is 
only linearly increasing with $L$ (rather than quadratically for a static medium).

Note finally that the condition $t_f\gg t_0$ that we used in deriving \eqn{omega0fin} is satisfied
for any point on the boundary $\omega=\omega_0(\theta)$ provided it is satisfied for the
largest allowed emission angle, equal to the jet angular opening $R$  (cf. Fig.~\ref{Fig:phase-space}).
Taking $\gamma=1$ for simplicity, this is equivalent to requiring
$R \ll 2$, which will be always satisfied in what follows.


\subsection{Colour decoherence in an expanding
  medium}\label{sec:decoherence-expanding}

To study the colour decoherence 
introduced by random collisions in the medium, it is customary to follow the propagation of a 
quark-antiquark antenna in a colour singlet state (a ``dipole'') and with a small opening angle
$\theta_0$~\cite{MehtarTani:2010ma,MehtarTani:2011tz,CasalderreySolana:2011rz,Mehtar-Tani:2014yea}.  Assume the antenna is created by a hard process occurring at $t=0$.
The probability for the antenna to remain a colour singlet  after crossing the medium
along a distance/time $t$ is $|S(t)|^2$, with $S(t)$ the  $S$-matrix for the elastic scattering
between the $q\bar q$ pair and the medium. Using the Gaussian approximation for
the dipole cross-section together with the fact that the transverse size of the dipole $r(t)$
grows linearly with time, $r(t)\simeq \theta_0 t$, one finds (see e.g.~\cite{CasalderreySolana:2011rz})
\begin{equation}\label{Sdip}
S(t)\simeq\exp\left(-\frac{\theta_0^2}{4} \int_{t_0}^{t}\dif t'\,\hat q (t')\,t^{\prime 2}\right)=
\exp\left(-\frac{\theta_0^2 \hat q(t)t^3}{4}\,\frac{1-(t_0/t)^{3-\gamma}}{3-\gamma}\right).
\end{equation}
The dipole has lost colour coherence, meaning that the quark and the antiquark act as
independent sources of radiation, when $|S|\ll 1$ or, alternatively, when the exponent in \eqn{Sdip}
becomes of order one. Assuming $t\gg t_0$,  one finds the following estimate
for the decoherence time 
\begin{equation}\label{tcoh}
t_{\rm coh}(\theta_0)
=\left(\frac{4}{\theta_0^2\hat q_0 t_0^\gamma}\right)^{1/(3-\gamma)}
= t_0 \left(\frac{2}{\theta_0}\right)^{2/(3-\gamma)},\end{equation}
where we have used $\hat q_0=1/t^3_0$ to obtain the second equality. So it is consistent to assume
$t_{\rm coh}\gg t_0$ so long as $\theta_0 \ll 2$. Using this estimate for $t_{\rm coh}$, one can
verify that:

{\sf (a) Colour decoherence plays no role during the formation of the in-medium vacuum-like
cascades.} In particular, it does not alter the angular ordering of the successive VLEs. The respective argument goes exactly as for a static medium~\cite{Caucal:2018dla}.\footnote{Here is a short version of the argument, for completeness:
consider an antenna with opening angle $\theta_0$ which initiates a vacuum-like gluon emission with energy $\omega_1$ and at a large angle $\theta_1>\theta_0$. Together, the conditions $\omega_1 >\omega_0(\theta_1)$ and $\theta_1>\theta_0$ imply $t_f=2/(\omega_1\theta_1^2) < 
t_{\rm coh}(\theta_0)$, meaning that the gluon is coherently emitted by the two legs of the antenna.
Hence, this would-be large-angle emission is in fact suppressed by interference effects, like
for antennas in the vacuum.}

{\sf (b) The decoherence time $t_{\rm coh}(\theta)$ becomes equal to the medium size $L$ when
$\theta=\theta_c(L)$, with $\theta_c(L)$ defined in \eqn{omegac}.} This means that antennas
with opening angles $\theta \gg \theta_c(L)$ rapidly lose colour coherence after formation 
and thus act as independent sources for MIEs, or for VLEs outside the medium. Vice-versa, 
antennas with small opening angles $\theta \le \theta_c(L)$ remain coherent throughout the medium, 
so they radiate MIEs as their parent partons would do.
This implies that emissions at small angles 
$\theta \le \theta_c(L)$ are not influenced by the medium
and hence can always be treated as {\it out-of-medium} emissions,
irrespective of their actual formation time (smaller or larger than $L$).
This explains the horizontal, lower, boundary, $\theta = \theta_c(L)$
delimiting the phase-space for in-medium VLEs in Fig.~\ref{Fig:phase-space}.

\subsection{Medium-induced emissions (MIEs) in a longitudinally
  expanding plasma}\label{MIEs-expanding}

As in the static case, our main purpose
is to provide a faithful description for the relatively soft emissions, with short formation times
$t_f\ll L$, for which multiple branching is potentially important. In a static medium, such emissions
can occur anywhere inside the medium, with a uniform rate. For the expanding medium, we expect
a bias towards the early time, when the medium is denser. Yet, we anticipate that the formation
times $t_f(\omega,t)$ for the soft emissions are still much smaller than the average time $t$
of their emission: $t_f(\omega,t)\ll t$. Under this assumption, that
we shall justify {\it a posteriori},
the soft emissions are quasi-local processes in time, which proceed as in the static medium,
except for the fact that their rate (or formation time) is controlled by the instantaneous value
$\hat q(t)$ of the jet quenching parameter at the (average) time of their formation. 
So, their (time-dependent) formation time can be estimated as 
\begin{equation}\label{tmed}
t_{f}(t)\,=\,\sqrt{\frac{2z(1-z)xE}{ \hat
    q_{\text{eff}}(t,z)}}\
   \overset{z\ll 1}{\approx}\ \sqrt{\frac{2\omega}{\hat q(t)}}\,,\qquad 
   \hat q_{\text{eff}}(t,z)\equiv\hat q(t) \big[1-z(1-z)\big],
\end{equation}
for the case of a $g\to gg$ splitting, where the parent gluon has 
energy $xE$ and the splitting fraction is equal to $z$. For $z\ll 1$, we used
$\omega\equiv zxE$ to denote the energy of the soft emitted gluon.\footnote{
One can  ``derive''
\eqn{tmed} by recalling that the transverse momentum $k_\perp$ of a 
MIE is acquired via collisions during formation, i.e.\ it is given by \eqn{ktbroad} with
$t_0\to t$ and $t\to t+t_f$. For $t_f\ll t$, this yields
$\langle k_\perp^2 \rangle (t+t_f,t)\simeq
\hat q(t) t_f$, which in turn implies $t_f=\frac{2\omega}{\hat q(t) t_f}
=\sqrt{\frac{2\omega}{\hat q(t)}}$ in agreement with~\eqref{tmed}. \label{FNtf}}
Then the time-dependent splitting rate follows as (with $\abar=\alpha_s N_c/\pi$)
\begin{equation}\label{BDMPSZrate}
 \frac{\rmd^2\Gamma_{\rm med}}{\rmd z \rmd t}=
 \frac{\alpha_sP_{g\to gg}(z)}{\sqrt{2}\pi}\frac{1}{t_{f}(t)}
 =\abar \frac{[1-z(1-z)]^2}{z(1-z)}\,
 \sqrt{\frac{\hat{q}_{\text{eff}}(t,z)}{z(1-z)xE}}\,.
\end{equation}
When used as a rate for successive gluon branchings in a Markovian
process, this expression is strictly valid only for emissions which
are soft enough for the condition $t_f(\omega,t)\ll t$ (or,
equivalently, $\omega\ll \hat q(t)t^2/2$) to be satisfied at generic
times $t\gg t_0$. We now show that this condition is indeed satisfied
for the soft gluons subjected to multiple branching. To that aim, we
first estimate the probability for emitting a single gluon with
energy  $\omega=zE\ll E$ up to time $t$. This is
obtained by integrating~\eqref{BDMPSZrate} up to a time $t$, which gives
(for $t\gg t_0$, and taking $x=1$)
\begin{equation}
\label{intrate}
 \int_{t_0}^t\rmd t'\,
 \frac{\rmd^2\Gamma_{\rm med}}{\rmd z \rmd t'}\simeq \abar\int_{t_0}^t\rmd t'
  \sqrt{\frac{\hat q(t')}{\omega}}\,\simeq\,\abar\,\frac{2}{2-\gamma}\,\sqrt{\frac {\hat q(t)t^2}
  {\omega}}\,.
  \end{equation}
 This  probability becomes of order one, meaning that multiple branching becomes important,
 when $\omega\lesssim\ombr(t)$, with
\begin{equation}\label{ombr}
 \ombr(t)\,\equiv\,\frac{4}{(2-\gamma)^2}\,\frac{\abar^2\hat
   q(t)t^2}{2}\,\ll\,\frac{\hat q(t)t^2}{2} \,.
\end{equation}
The strong inequality above, valid in the weak coupling regime at
$\abar^2\ll 1$, confirms that the condition $t_f(\omega,t)\ll t$ is
well satisfied for the soft gluon emissions with
$\omega\lesssim\ombr(t)$.
 
That said, the rate in \eqn{BDMPSZrate} can also be used for harder
emissions, which are rare (i.e.\ which occur at most once over a time
$t\sim L$), provided its time integral up to $L$ reproduces the
expected result for the BDMPSZ spectrum in an expanding
medium~\cite{Baier:1998yf,Zakharov:1998wq,Arnold:2008iy}. This is
indeed the case as one can see by by replacing $t\to L$ in
\eqn{intrate}, which gives
\begin{equation}\label{spec}
  \omega \frac{\rmd\mathcal{P}_{\rm med}}{\rmd \omega}
  = \int_{t_0}^L\rmd t\,
  \frac{\rmd^2\Gamma_{\rm med}}{\rmd z \rmd t}
  \,\equiv \,\abar\,\sqrt{\frac {2\tilde\omega_c(L)}
  {\omega}}\,\overset{L\gg t_0}{\simeq}\,\abar\,\frac{2}{2-\gamma}\,\sqrt{\frac {\hat q(L) L^2}
  {\omega}},
\end{equation}
where we introduced (recall the definition of $\omega_c(L)$ in \eqn{omegac})
\begin{equation}\label{eq:omegactilde}
  \tilde\omega_c(L)
  \equiv
  \,\frac{2}{(2-\gamma)^2}\,\qhat(L)L^2\left[1-(t_0/L)^{1-\frac{\gamma}{2}}\right]^2
  \overset{L\gg t_0}{\simeq} \frac{4}{(2-\gamma)^2}\,\omega_c(L).
\end{equation}
Eq.~(\ref{spec}) is the right limit  of
the general result~\cite{Baier:1998yf,Zakharov:1998wq,Arnold:2008iy}
for sufficiently low energies $\omega  \ll \tilde\omega_c(L)$
and remains a good approximation (at least, parametrically) up to
$\omega  \sim \tilde\omega_c(L)$.

The upper limit $\tilde\omega_c(L)$ is  natural in this context,
as this is (parametrically) the energy of a MIE  with formation time $t_f\sim L$. And indeed,
the full  BDMPSZ  spectrum in Refs.~\cite{Baier:1998yf,Zakharov:1998wq,Arnold:2008iy} is rapidly
decreasing for larger energies  $\omega  \gg \tilde\omega_c$, like the power
$(\tilde\omega_c/\omega)^2$. To mimic that, we simply supplement our simplified rate
in  \eqn{BDMPSZrate} with a sharp cutoff at $\omega = \tilde\omega_c(L)$. This approximation
has essentially no impact on the observables that we are primarily interested 
in,\footnote{We intend to relax this approximation in a future work, 
 by using a branching rate derived
from the complete expression for the BDMPSZ spectrum, as done
e.g. in Refs.~\cite{Adhya:2019qse}. For the time being, we have performed 
numerical tests using an approximation for the rate which interpolates between the right
behaviours at both small and large energies (compared to $\tilde\omega_c(L)$)
and found only negligible effects on quantities like the $R_{AA}$ ratio.} like
the nuclear modification factor $R_{AA}$,  which are controlled
by relatively soft emissions with $\omega  \ll \tilde\omega_c(L)$. 

For later convenience, we note that if one introduces the ``dimensionless time''
$\tau$ such that
\begin{equation}\label{reduced-time-exp}
 \frac{\dif\tau}{\dif t}\equiv \sqrt{\frac{\qhat(t)}{E}}\ \Longrightarrow \ 
 \tau(t,t_0)=\int_{t_0}^t\dif t'\,\sqrt{\frac{\qhat(t')}{E}}\,=
 \,\frac{2}{2-\gamma}\,\sqrt{\frac {\hat q(t)t^2}
  {E}}\,\left[1-\left({t_0}/{t}\right)^{1-\frac{\gamma}{2}}\right]\,,
\end{equation}
the rate for MIEs, Eq.~(\ref{BDMPSZrate}), becomes time-independent when rewritten
as a rate in $\tau$:
\begin{equation}\label{BDMPSZrate-dimensionless}
 \frac{\rmd^2\Gamma_{\rm med}}{\rmd z \rmd \tau}= 
\abar \frac{\left[1-z(1-z)\right]^{5/2}}{\sqrt{x}\left[z(1-z)\right]^{3/2}}\,.
\end{equation}

We conclude this section with a comment on the value of the QCD coupling to be used for MIEs.
In all previous formul\ae, we treated this as a {\it fixed} coupling.
Physically though,  one should use the QCD running coupling at a scale of the order of the
relative transverse momentum of the emitted gluon, as acquired during the formation time.
This scale depends upon the kinematics of the emission and, for an expanding medium, also upon
{\it time}: $\alpha_s(k_f^2)$ with $k_f^2\simeq \hat q(t) t_f\simeq\sqrt{2\hat q(t)\omega}$ (cf.\ footnote~\ref{FNtf}).
That would of course change most of the previous formul\ae\ in this
section (in particular, the definition~\eqref{reduced-time-exp} of the
reduced time).
In what follows, we shall however keep the fixed-coupling approximation for 
the MIEs, like we have
done in our previous studies (see e.g.~\cite{Caucal:2019uvr}), and
postpone the inclusion of the respective running-coupling corrections  to a
future work.

\subsection{Transverse momentum broadening in an expanding
  plasma}\label{sec:broadening-expanding}

As already mentioned, the partons propagating through the plasma suffer transverse momentum broadening 
via multiple soft scattering, with a rate equal to  $\hat q(t)$. In this section, we shall study this process in
more detail and establish some results that will be useful later, when interpreting our Monte-Carlo predictions.

For the partons created via VLEs, the transverse momentum acquired via elastic collisions adds to that
generated at the emission vertex.  For those produced via MIEs, the collisions represent the only
source of transverse momentum.
The collisions can occur both during the 
quantum emission process, in which case they determine the formation time $t_f(\omega)$ according to
\eqn{tmed}, and during the propagation of the daughter partons until they split again, or until they
leave the medium. For the soft emissions, with $\omega\ll \tilde\omega_c(L)$,
to which the rate \eqref{BDMPSZrate} strictly applies, the formation time is comparatively small,
$t_f(\omega)\ll L$, and one can neglect the transverse momentum broadening during formation: that is,
one can treat a MIE as a {\it collinear} branching.
After formation, the daughter partons suffer elastic collisions, leading to a random walk in 
transverse momentum space.

For a parton created at time $t_1$ and which decays at time $t_2\le L$
(with $t_2= L$ for a parton exiting the medium before splitting),
its transverse momentum with respect to its parent parton is sampled according to the Gaussian distribution,
\begin{equation}\label{2d-broadening}
 \frac{\dif^2 \mathcal{P}_{\rm broad}}{\dif^2 k_\perp}=\frac{1}{\pi\langle k_\perp^2\rangle(t_2,t_1)}\exp\left(\frac{-k_\perp^2}{\langle k_\perp^2\rangle(t_2,t_1)}\right),
\end{equation}
with  the average $\langle k_\perp^2\rangle(t_2,t_1)$ given by \eqn{ktbroad}.

In the Monte Carlo simulations, we shall use this probability distribution to generate
the transverse momentum acquired by each parton from the moment $t_1$ 
when it is created in the plasma
until the moment $t_2$  when it decays again, or it leaves the medium. However, for the sake of 
analytic arguments to be discussed in Sect.~\ref{sec:scaling}, we would also need a more general
estimate for the average transverse momentum squared $\langle k_\perp^2\rangle(t_2,t_1)$, in which one  averages over the initial and final times ($t_1$ and $t_2$, respectively).
For relatively large energies $\om\gg \ombr$, where 
multiple branching is negligible, such an estimate is easy to obtain: an energetic parton
propagates through the medium along a distance $\sim L$ and thus acquires a typical 
$k_\perp$--broadening given by \eqn{ktbroad} with $t\to L$.

The corresponding calculation for the softer gluons with energies
$\omega\lesssim \omega_\text{br}$ is more subtle as one must take the
multiple branchings into account. In the case of a static medium, a simple estimate
can be found via the following, intuitive, argument:  gluons with
$\om\lesssim \ombr$ have a typical lifetime (from their emission to
their splitting) $\tbr\sim (1/\abar)t_f$, with
$t_f=\sqrt{2\omega/\hat q}$. During
this lifetime, the emission will accumulate a transverse momentum
broadening $\bar{k}_\perp^{\,2}(\omega)\simeq \hat q \tbr\sim
(1/\abar)\sqrt{\omega\hat q}$. 

A more precise calculation\footnote{This calculation
  involves some simplifying assumptions, notably a simpler version for
  the branching kernel (see Appendix~\ref{app:A} for details), which
  do not alter the qualitative features that we are presently
  interested in.} presented in~\cite{Blaizot:2014ula}  has confirmed this simple
  estimate and also fixed the overall proportionality coefficient:
   $\bar{k}_\perp^{\,2}(\omega)\simeq\sqrt{\omega\hat q}/4\abar$ for
$\om\ll\ombr$  (see also the related studies in~\cite{Blaizot:2014rla,Kutak:2018dim,Rohrmoser:2020ltf,Blanco:2020uzy,Barata:2020rdn}).
In  Appendix~\ref{app:A}, we generalise the calculation of Ref.~\cite{Blaizot:2014ula}
to the case of an expanding medium. Interestingly, we find
$\bar{k}_\perp^{\,2}(\omega)\simeq\sqrt{\omega\hat q(L)}/4\abar$, which is formally identical
to that for the static case up to the replacement $\hat q\to \hat q(L)$.
The appearance of the late-time quenching parameter $\hat{q}(L)$ comes
from the fact that very soft gluons ($\om\lesssim \ombr$) are predominantly
created at the latest stages of the jet evolution, when the number of
their sources (other soft gluons) is largest.

\section{Monte-Carlo implementation and choice of
  parameters}\label{sec:mc}

Most of the picture described in Sect.~\ref{sec:ps} can be easily
implemented in a parton-shower Monte Carlo program. For the case of a static
medium, this implementation has been presented in great detail in Ref.~\cite{Caucal:2019uvr}.
In what follows, we shall discuss only the specific adjustments which are needed
in order to take the expansion of the medium into account.

As far as the VLEs are concerned, the only modification associated
with the expansion of the medium is the different kinematic
boundary for the VLEs emitted inside the medium (i.e.\ in the red
region of Fig.~\ref{Fig:phase-space}). For this, we directly use
Eq.~(\ref{omega0fin}) with an endpoint given by Eq.~(\ref{omegac}).
The branching probability and the running-coupling prescription
for the QCD coupling at the emission vertex are as presented 
in Ref.~\cite{Caucal:2019uvr}.

Medium-induced emissions are obtained by implementing the
dimensionless-time rate (\ref{BDMPSZrate-dimensionless})
together with its generalisations to other
partonic channels,\footnote{The Casimir factor and the $z$--dependence
  depend on the decay channel of the splitting;
  see~\cite{Mehtar-Tani:2018zba} for explicit expressions.}
with $\tau$ defined by Eq.~(\ref{reduced-time-exp}). As in Ref.~\cite{Caucal:2019uvr},
we evaluate this rate with a fixed value for  the QCD coupling, 
that we denote as $\alpha_{s,\rm med}$ and will be treated as a free parameter.
The use of a more physical prescription for this coupling, along the
lines discussed at the end of Sect.~\ref{MIEs-expanding}, should be one of the
objectives of a future upgrade of our MC implementation.

Still concerning the MIEs, we impose an upper limit on their energies in the form of a 
sharp cutoff at $\omega=\tilde\omega_c$, cf.\ Eq.~(\ref{eq:omegactilde}). The
transverse momentum broadening is generated according to the Gaussian
distribution, Eq.~(\ref{2d-broadening}), where the initial and final
physical times $t_1$ and $t_2$ are obtained from their dimensionless
equivalents, $\tau_1$ and $\tau_2$, by inverting
explicitly~(\ref{reduced-time-exp}).

We note the small mismatch between the scale $\tilde\omega_c$, used as
the endpoint for the energy of medium-induced emissions, and
$\omega_c$, used to define the phase-space available for VLEs.
In the leading, double-logarithmic, accuracy to which we control
the boundaries of the phase-space available for VLEs, we could equally use
$\omega_c$ or $\tilde\omega_c$ to define the lower endpoint of the vetoed region
(recall the discussion after \eqn{omega0}).
 In practice, we have decided to use $\omega_c$ which not only is simpler,
but also most naturally follows from our physical arguments in 
Sect.~\ref{sec:VLEs-in-expanding}.

\begin{table}
  \centering
  \begin{tabular}{c|ccccc|ccc|c}
    $Q_s$ & $t_0$ & $L$ & $\hat{q}_0$ & $\hat{q}(L)$ & $\amed$
    & $\tilde\omega_c$ & $\omega_\text{br}$ & $\theta_c$  &  $\hat{q}_\text{stat}$ \\
    {\footnotesize [GeV]} & {\footnotesize [fm]} & {\footnotesize [fm]}
     & {\scriptsize [GeV$^2$/fm]} & {\scriptsize [GeV$^2$/fm]} & 
    & {\footnotesize [GeV]} & {\footnotesize [GeV]} & & {\scriptsize [GeV$^2$/fm]} \\
    \hline
    1.2 & 0.1667 & 4 &  8.64 & 0.36 & 0.35 & 36.48 & 4.08 & 0.0833 & 0.99 \\
    1.4 & 0.1429 & 4 & 13.72 & 0.49 & 0.28 & 51.57 & 3.68 & 0.0714 & 1.39 \\
    {\bf 1.6} & {\bf 0.125}  & {\bf 4} & {\bf 20.48}
        & {\bf 0.64} & {\bf 0.23}
    & {\bf 69.40 } & {\bf 3.35 } & {\bf 0.0625}  & {\bf 1.85} \\
    2   & 0.1    & 4 & 40    & 1.0  & 0.17 & 113.4 & 2.99 & 0.05 & 2.98 \\
  \end{tabular}
  \caption{Set of parameters that we consider throughout this
    paper. The values of $\amed$ are adjusted so as to obtain a good
    description of the $R_{AA}$ jet nuclear suppression factor (see
    section~\ref{sec:pheno}). The row highlighted in bold corresponds to our
    default set.}\label{tab:parameters}
\end{table}

As described in Ref.~\cite{Caucal:2019uvr}, our simple (collinear)
implementation for the vacuum part of the parton shower requires the
introduction of a maximal angle, that we set here to
$\theta_\text{max}=1$, and of a minimal $k_\perp$ for vacuum
emissions, that we set here to
$k_{\perp,\text{min}}=0.25$~GeV.\footnote{In practice, these scales
  can be varied to gauge the size of the uncertainties associated with
  our simple modelling of vacuum parton showers, as we have done in
  Refs.~\cite{Caucal:2019uvr,Caucal:2020xad}. For simplicity, we do
  not do this here.}
We then have to fix the parameters describing the expanding medium.
In the numerical calculations to follow, we will consider $\gamma=1$ for an expanding
medium and compare our results with a static medium ($\gamma=0$). We
will fix the medium length to $L=4$~fm.
The two parameters $t_0$ and $\hat{q}_0$ characterising
the expansion of the medium are both fixed in terms of the initial saturation momentum
$Q_s$, using $t_0=1/Q_s$ and $\hat{q}_0t_0=Q_s^2$. The different sets of medium
parameters we study are listed in table~\ref{tab:parameters}. For each value of
$Q_s$, the coupling $\amed$ has been roughly adjusted so as to provide
a good description of the LHC data~\cite{Aaboud:2018twu} for the jet
nuclear modification factor $R_{AA}$ (see section~\ref{sec:RAA}). For
completeness, we list in table~\ref{tab:parameters} the parameters
$\tilde\omega_c$ from Eq.~(\ref{eq:omegactilde}) as well as
$\omega_\text{br}=\baramed^2\tilde\omega_c$, the typical scale at which
multiple branching becomes important for MIEs, and the decoherence
angle $\theta_c$, Eq.~(\ref{omegac}).
The last column gives the ``static-equivalent'' value of the quenching
parameter $\hat{q}$. It is defined in Eq.~(\ref{qeff}) below, a choice that
we shall discuss at length in section~\ref{sec:scaling}.
The set of parameters shown in bold characters in table~\ref{tab:parameters}
corresponds to our default choice.

\section{Scaling properties of jet fragmentation: expanding vs. static media}
\label{sec:scaling}

In this section, we highlight an interesting scaling property between
the longitudinal spectrum of MIEs in an expanding medium and that of
a suitably-defined static medium.
We first derive this scaling in section~\ref{sec:scaling-mies}, then
discuss two sources of scaling violations: \texttt{(i)} the transverse momentum broadening
and its impact, first, on the jet fragmentation function
(section~\ref{sec:scaling-violations-broadening}) and, second, on the energy loss via
MIEs by a single hard parton
(section~\ref{sec:eloss-parton}), and \texttt{(ii)}  VLEs
emitted inside the medium (section~\ref{sub:full-show}).

\subsection{Scaling and the static-equivalent medium}
\label{sec:scaling-mies}

After introducing the dimensionless time variable $\tau$ in
Eq.~(\ref{reduced-time-exp}), the rate
$ \frac{\rmd^2\Gamma_{\rm med}}{\rmd z \rmd \tau}$ for MIEs in the expanding
medium is independent of $\tau$ and formally identical to that of a
static medium, for which the $\tau$-variable is naturally defined as
$\tau(t)\equiv t\sqrt{\hat q/E}$.
Since the rates in $\tau$ are identical, so are the respective energy
distributions for the partons produced via MIEs after a (physical)
time $L$, provided the jet quenching parameter $\hat q_{\rm stat}$ for
the ``equivalent'' static problem satisfies\footnote{We choose the
  size $L_{\rm stat}=L-t_0$ of the equivalent static problem to be the
  same as that of the expanding medium. This is convenient when
  discussing the physical correspondence between the 2 problems, but
  it is not required at a mathematical level as \eqn{qeff} only
  constrains the product $\hat q_{\rm stat}L_{\rm stat}^2$.}
$\tau_{\rm stat}(L-t_0)=\tau(L,t_0)$, i.e.
\begin{equation}\label{qeff}
\qhat_{\rm stat} =\left(\frac{1}{L-t_0}\int_{t_0}^L\dif t'\,\sqrt{\qhat(t')}\right)^2
=\frac{4}{(2-\gamma)^2}\,\qhat(L)\left(\frac{1-(t_0/L)^{1-\frac{\gamma}{2}}}{1-t_0/L}\right)^2
\overset{L\gg t_0}{\simeq} \,\frac{4}{(2-\gamma)^2}\,\hat q(L)\,.
\end{equation}
For this equivalence to hold, we also need to make sure that this
static-equivalent medium has the same upper limit on the energy
spectrum, namely $\omega=\tilde\omega_c$, cf. Eq.~(\ref{eq:omegactilde}).
This cutoff can be equivalently written as $\tilde\omega_c=\qhat_{\rm stat} (L-t_0)^2/2$,
which is precisely the form of the corresponding cutoff for a static medium,
as used in our previous studies~\cite{Caucal:2018dla,Caucal:2019uvr,Caucal:2020xad} .

In particular, the spectrum~\eqref{spec} for an expanding medium is formally
identical to that of a static medium $ \omega \frac{\rmd\mathcal{P}_{\rm stat}}{\rmd \omega}
=\abar\sqrt{\frac{2\tilde \omega_c}{\omega}}$.
This is also the case for the characteristic energy scale $\ombr$ for the onset of multiple
branching since, for $t=L$, the estimate in \eqn{ombr} coincides with
the corresponding static-equivalent scale
 $\ombr=\abar^2\qhat_{\rm stat} (L-t_0)^2/2$.

Note that the full dependence on $t_0/L$ in~(\ref{qeff}) has to be
kept for the exact scaling to be satisfied. This is what we do for the
simulation results presented in this section.\footnote{In practice,
  the exact $\hat{q}_\text{stat}$ can be up to 30\% smaller than its
  asymptotic value when $L/t_0\to\infty$
  (which for $\gamma=1$ is equal to $4\hat{q}(L)$), as visible in
  table~\ref{tab:parameters}.} That said, in physical considerations and
  parametric estimates, we shall often neglect $t_0$ next to $L$, for simplicity; e.g.,
  we shall simply write $\ombr=\abar^2\qhat_{\rm stat} L^2/2$ for qualitative purposes.
  
  To avoid potential confusion, it is useful
to stress that the scaling law~\eqref{qeff}, which refers to the
rate~\eqref{BDMPSZrate} for relatively soft ($\omega  \ll \tilde\omega_c$) gluon emissions, is not the
same as the  original scaling law identified in~\cite{Baier:1998yf}, which instead refers to the
average energy loss by the leading parton, $\Delta E\equiv \int\rmd\omega  \,
\omega \frac{\rmd\mathcal{P}_{\rm med}}{\rmd \omega}$. This quantity $\Delta E$ is controlled
by the most energetic MIEs, with energy $\omega\sim  \tilde\omega_c$, which are not accurately
described by our approximate spectrum~\eqref{spec}. The correct calculation of 
$\Delta E$ for the expanding medium in Ref.~\cite{Baier:1998yf} yields a different scaling law:
$\Delta E_{\rm exp}=\Delta E_{\rm static}$, with $\Delta E_{\rm static}$ computed with 
$\qhat_{\rm stat} \to \langle \hat q\rangle$, where $\langle \hat
q\rangle$ is the following time average of $\hat q(t)$:
\begin{equation}\label{aveq}
  \langle \hat q\rangle\equiv \frac{2}{(L-t_0)^2} \int_{t_0}^L\rmd
  t\,t\,\hat q(t)
  \overset{L\gg t_0}\simeq\frac{2}{2-\gamma}\, \hat q(L)\,.
\end{equation}
This $\langle \hat
q\rangle$ is different and actually smaller (for any  $\gamma >0$) than our $\qhat_{\rm stat} $ in 
\eqn{qeff}; e.g.\ $\qhat_{\rm stat} \simeq 2 \langle \hat q\rangle$ for $\gamma=1$. A recent 
numerical study~\cite{Adhya:2019qse} of the scaling properties of the full BDMPSZ spectrum 
for an expanding medium shows that the scaling law \eqref{qeff} is indeed well satisfied at low 
energies $\omega  \ll \tilde\omega_c$, whereas that in \eqn{aveq} becomes the correct scaling
at larger energies $\omega  \sim \tilde\omega_c$. (See also Refs.~\cite{Salgado:2002cd,Salgado:2003gb}
for previous numerical studies of the quality of the scaling law in \eqn{aveq}.)

Within our present description of MIEs, where the ``soft'' emission
rate~\eqref{BDMPSZrate} is used for all energies $\omega$ up
$\tilde\omega_c$, the scaling law~\eqref{qeff} is exactly
satisfied for the energy distributions produced via MIEs alone.
In a more generic context, this scaling is violated by two main
effects: transverse momentum broadening and VLEs occurring inside the
medium.
Thus, we expect deviations from the scaling laws in the full (energy
and angle) parton distributions, produced via both VLEs and MIEs.
We study the quality of these scaling properties via Monte Carlo
simulations and analytic estimates in the remaining part of this section.

\subsection{Scaling violations from transverse momentum broadening}\label{sec:scaling-violations-broadening}

Transverse momentum broadening introduces violations of the scaling
properties discussed in section~\ref{sec:scaling-mies} since its rate
$\rmd \langle k_\perp^2\rangle/\rmd t=\hat q(t)$ scales linearly in
  $\hat{q}(t)$, unlike  the emission rate \eqref{BDMPSZrate}, which scales
  like $\sqrt{\hat q(t)}$.
In principle, exact scaling can be recovered by integrating out the
parton transverse momenta to obtain inclusive energy
correlations.
In practice however, when measuring energy correlations in a jet
(e.g.\ the fragmentation function), one excludes from the
integration over transverse momenta those propagating outside the jet,
i.e.\ at angles $\theta\simeq k_\perp/\omega$ larger than the jet
radius $R$. This induces a violation of scaling, even for 
inclusive energy distributions.

To study numerically the scaling violations associated
with $k_\perp$--broadening, we compute the fragmentation function
$D(\omega)\equiv \omega(\rmd N/\rmd\omega)$ for events generated via
MIEs alone, starting with a leading gluon with energy
$p_{T0}=200$~GeV. The resulting partons are clustered with the
anti-$k_t$ algorithm with a radius $R$, keeping the hardest resulting jet.

\begin{figure}[t] 
  \centering
  \begin{subfigure}[t]{0.48\textwidth}
    \includegraphics[width=\textwidth]{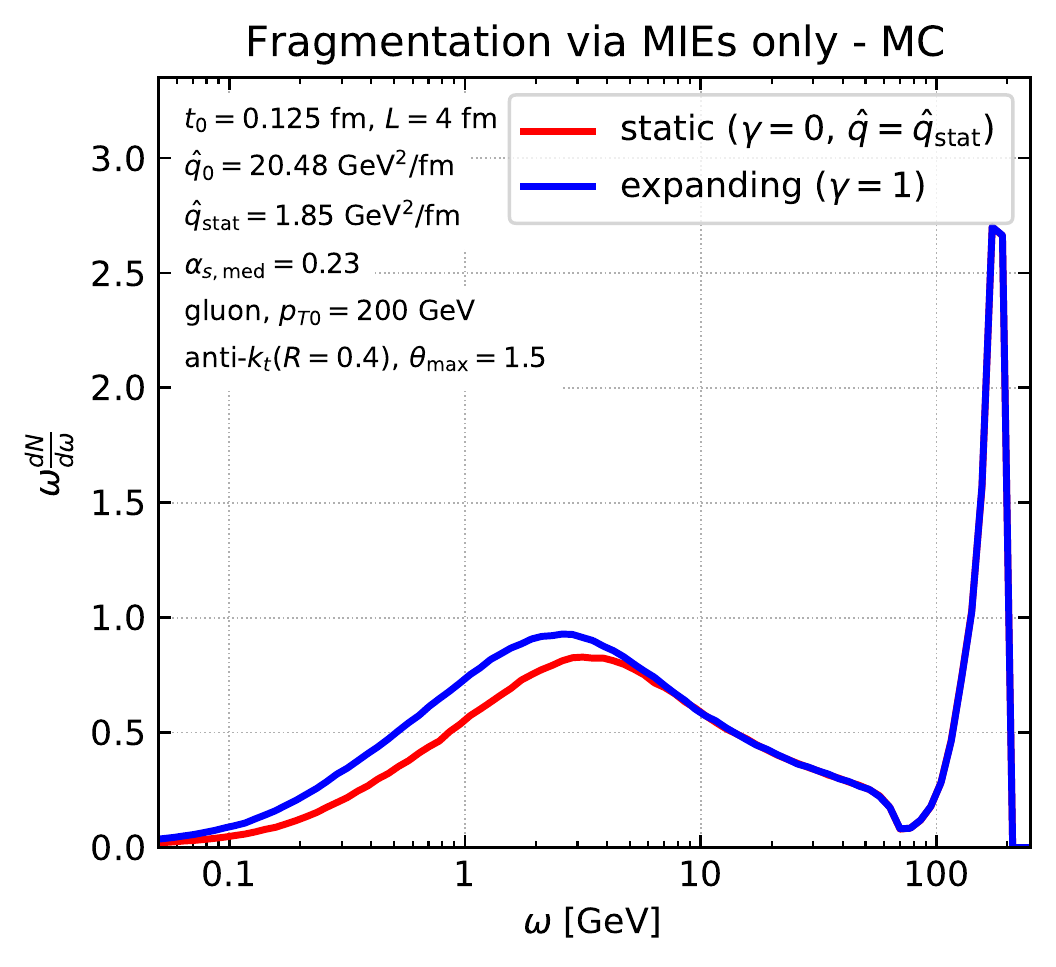}
    \caption{\small Monte-Carlo results for the parton distribution inside a gluon-jet generated via MIEs alone in a Bjorken expanding medium (blue) and its equivalent static medium (red).}\label{Fig:frag-func-1}
  \end{subfigure}
  \hfill
  \begin{subfigure}[t]{0.48\textwidth}
    \includegraphics[width=\textwidth]{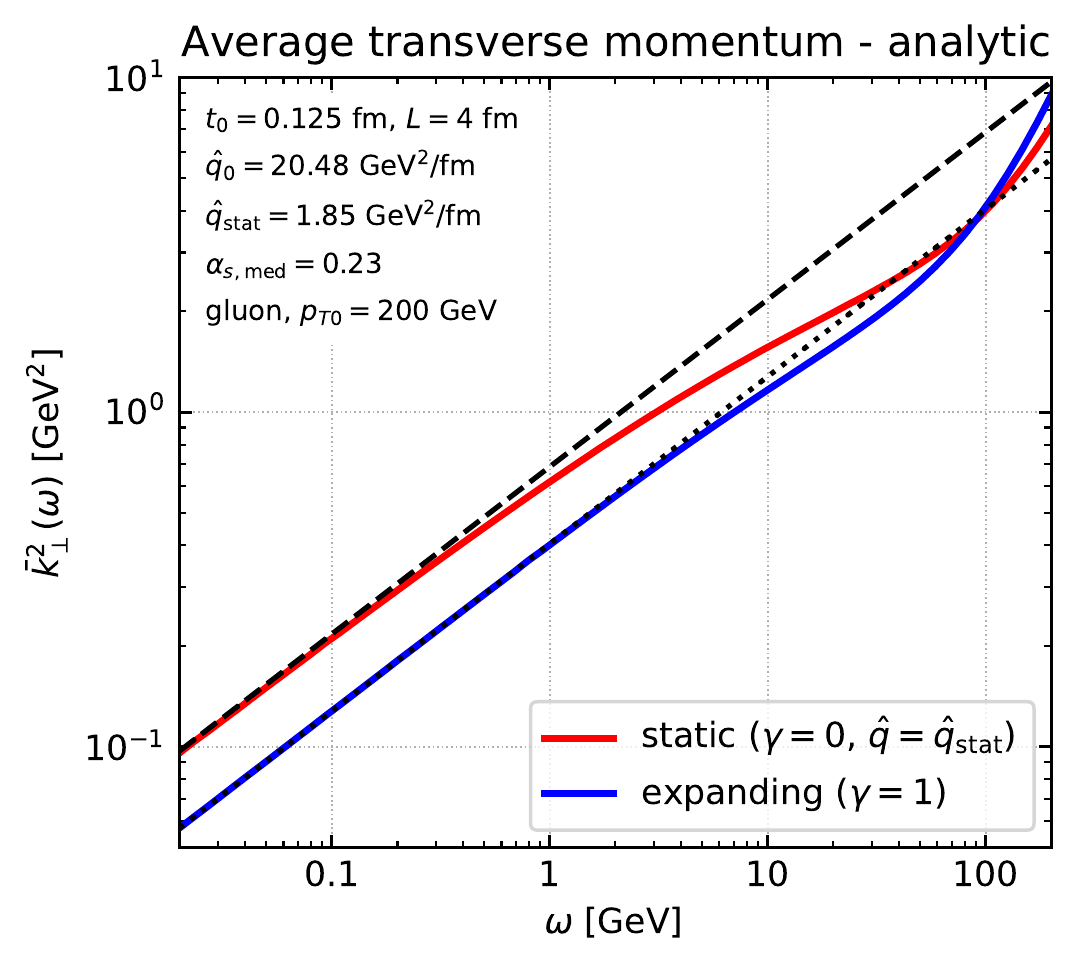}
    \caption{\small Average transverse momentum squared acquired by a parton in a
    (purely gluonic)
     medium induced cascade, in two scenarios:  expanding medium (blue) and equivalent static medium (red). The curves are obtained by numerically evaluating Eq.~\eqref{W-sol}.}\label{Fig:frag-func-2}
  \end{subfigure}
  \caption{\small Scaling violations induced by transverse momentum
    broadening in medium-induced cascades.  For all these results, the jet is triggered by a leading gluon 
    with energy $p_{T0}=200$ GeV.
  }\label{Fig:frag-func} 
\end{figure}

Fig.~\ref{Fig:frag-func-1} shows this distribution for the expanding
medium with $\gamma=1$ (the blue curve) and for
the ``equivalent'' static medium (the red curve labelled $\gamma=0$),
for our default set of parameters (the boldface line in table~\ref{tab:parameters}).
One sees that for sufficiently high energies 
$\omega \gtrsim 5$~GeV, the static and expanding results are
indistinguishable from each other, indicating perfect scaling.
The narrow peak at $\omega \sim  p_{T0}$ represents the leading gluon, 
the minimum at $\om \sim \tilde\omega_c$ corresponds to the upper
bound on the radiation spectrum, and the increase with decreasing $\om$ below 
$\tilde\omega_c$ is the expected growth\footnote{\eqn{spec} is
the spectrum in the single emission approximation, but the $1/\sqrt{\om}$ behaviour
at low energies is preserved by multiple branching~\cite{Baier:2000sb,Blaizot:2013hx}.}
 $\propto 1/\sqrt{\om}$, cf.\ \eqn{spec}.
However, the scaling is broken at lower energies $\omega \lesssim 5$~GeV,
where both distributions show a broad peak.

These features are easy to understand.
Partons with large $\omega$
remain inside the jet even after transverse momentum broadening, hence their
energy spectrum is independent of the jet radius $R$ and scaling is
obeyed.
On the contrary, momentum broadening can deflect softer gluons to angles
$\theta\simeq k_\perp/\omega$ larger than $R$. This explains both the
decrease of $D(\omega)$ at low energies and the scaling violations, as we now explain.

To that aim, we rely on the analysis of transverse momentum broadening
in section~\ref{sec:broadening-expanding}. 
The broad peak visible in Fig.~\ref{Fig:frag-func-1} at intermediate energies corresponds
to the softest gluons whose propagation angle is still inside the jet:
$\theta\simeq k_\perp/\omega\lesssim R$, or $\omega \gtrsim k_\perp/R$. From Fig.~\ref{Fig:frag-func-1}, 
one sees that the relevant energies are comparable to
the medium scale $\ombr\simeq 3.35$~GeV for multiple branching, cf. 
 table~\ref{tab:parameters}.  For these gluons, the average transverse momentum
 $\bar{k}_\perp^{\,2}(\omega)$ is controlled by $\qhat(L)$ for the expanding medium, and by 
  $\qhat_{\rm stat}$ for the ``equivalent'' static one (see the discussion towards
the end of Sect.~\ref{sec:broadening-expanding} and in  Appendix~\ref{app:A}).
Since $\qhat(L)<\qhat_{\rm stat}$,
$\bar{k}_\perp^{\,2}(\omega)$ is smaller (for a given energy $\omega\lesssim\ombr$)
in an expanding medium than in the equivalent static one.
This is illustrated by the plot in Fig.~\ref{Fig:frag-func-2}, which shows numerical results for
$\bar{k}_\perp^{\,2}(\omega)$ for the two scenarios, as obtained via the method
outlined in Appendix~\ref{app:A} . 
Because of that, the soft gluons are more
likely to remain within the jet cone in an expanding-medium than in the
corresponding static one. In particular, the position of the low-energy peak in the spectrum
can be roughly estimated as $\bar\omega\simeq
\bar{k}_\perp(\bar{\omega})/R$. This implicit equation gives a result
$\bar{\om}  \simeq (\qhat /16\bar{\alpha}_s^2 R^4)^{1/3}$ which
is smaller for the expanding medium than for the static one (because 
$\hat{q}(L)<\hat{q}_\text{stat}$), in agreement with Fig.~\ref{Fig:frag-func-1}.

\subsection{Energy loss by the leading parton via MIEs} \label{sec:eloss-parton}

For phenomenological applications, it is interesting to understand the
average energy, $\varepsilon(p_{T0},R)$, lost by a jet initiated by a
hard parton of momentum $p_{T0}$ outside a cone of opening angle
$R$.
We cover the case with only MIEs in this section and discuss the case
of a full parton shower, including both VLEs and MIEs, in the next
section.

Within our effective theory,
 $\varepsilon(p_{T0},R)$ is the sum of two components~\cite{Blaizot:2013hx,Fister:2014zxa}: 
 \texttt{(i)} the energy 
flowing down to arbitrarily soft energies (hence, moving out to arbitrarily large angles), via multiple branchings;
this corresponds to the ``turbulent flow'' in the language of Refs.~\cite{Blaizot:2013hx,Fister:2014zxa}
and gives a contribution proportional to $\ombr$, and  \texttt{(ii)} 
the energy carried by primary emissions which are soft enough to
propagate at angles larger than $R$, i.e.\ gluons with energies $\om\lesssim\bar{\om}$, with $\bar{\om} 
  \simeq \big(\qhat(L)/16\bar{\alpha}_s^2 R^4\big)^{1/3}$ 
the scale introduced at the end of section~\ref{sec:scaling-violations-broadening}. Altogether, we can write
\begin{equation}\label{eq-eloss}
 \varepsilon(p_{T0},R)\simeq\upsilon\ombr+\int_0^{\bar{\om}}\dif \om\, \omega \frac{\rmd N}{\rmd \omega}
\end{equation}
with $\upsilon$ a constant.\footnote{$\upsilon\simeq 4.96$ for $p_{T0}<\om_c$~\cite{Baier:2000sb,Blaizot:2013hx}, whereas in the high-energy limit $p_{T0} \gg \om_c$, one finds $\upsilon\simeq 3.8$ for $\abar=0.24$~\cite{Fister:2014zxa}.} 
While the first component, inclusive in $k_\perp$, satisfies the
scaling behaviour w.r.t.\ the equivalent static medium,
the second component breaks the scaling through the upper limit
$\bar{\om}$ which comes from the transverse momentum broadening
(see section~\ref{sec:scaling-violations-broadening}).
Since $\bar{\om}_{\rm exp} < \bar{\om}_{\rm stat}$ (with obvious
notations), the partonic energy loss $\varepsilon(p_{T0},R)$ is
expected to be smaller in the expanding scenario. This is another consequence
of the fact that soft emissions are more likely to remain inside the
jet in an expanding medium.

\begin{figure}[t] 
  \centering
  \begin{subfigure}[t]{0.48\textwidth}
    \includegraphics[width=\textwidth,page=1]{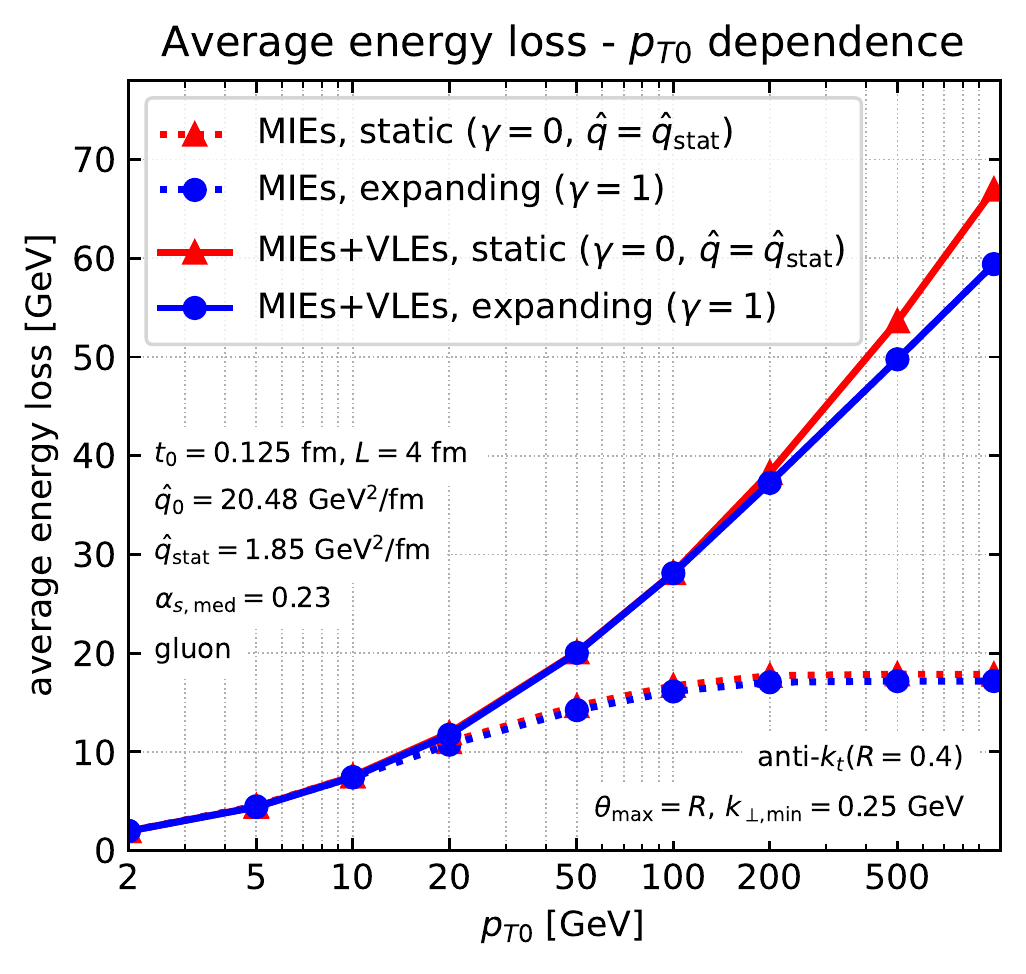}
    \caption{\small}\label{Fig:eloss-1}
  \end{subfigure}
  \hfill
  \begin{subfigure}[t]{0.48\textwidth}
    \includegraphics[width=\textwidth,page=2]{eloss-v-pt-expansion.pdf}
    \caption{\small }\label{Fig:eloss-2}
  \end{subfigure}
  \caption{\small Our MC results for the average energy loss by a
    gluon-initiated jet are displayed as a function of the initial
    energy $p_{T0}$ of the leading parton (left) and the jet radius
    $R$ (right), for two scenarios for the jet evolution: jets with
    MIEs only (dotted lines) and full showers with both MIEs and VLEs
    (plain lines). For each scenario, we compare the case of an expanding medium
    (blue) with the equivalent static one (red).}\label{Fig:eloss}
\end{figure}

To verify this and to check the functional dependence
predicted by  \eqn{eq-eloss}, we compute the average energy loss $\varepsilon(p_{T0},R)$ for
a gluon initiated jet and for various values of
the initial energy  $p_{T0}$ and the jet radius $R$, using again our default
set of medium parameters.
The results are shown as dotted lines in Fig.~\ref{Fig:eloss}, for
both the expanding medium (blue curves) and the equivalent static one
(blue curves). While they confirm that
the energy loss is slightly smaller for an expanding medium, the
differences are barely visible, corresponding
to an almost perfect scaling. 

To better understand this, we consider the scaling violations induced
by the upper limit $\bar{\om}$ in the second term in \eqn{eq-eloss}.
Using $\omega(\rmd N/\rmd\omega) \propto 1/\sqrt{\om}$,
cf.\ \eqn{spec}, one sees that this term scales like
$\bar{\om}^{1/2}\propto \qhat^{1/6}$. The contribution from this
second term will therefore be larger for the static medium only by a
moderate factor of
$(\qhat_{\rm stat}/\qhat(L))^{1/6}\simeq 4^{1/6}\simeq 1.26$.

The dotted lines in Fig.~\ref{Fig:eloss-2} show the dependence of
$\varepsilon(p_{T0},R)$ on $R$ (for fixed $p_{T0}= 200$~GeV), for the two
scenarios for the medium.
This  dependence comes from the upper limit
$\bar{\om}$ in the integral term of Eq.~(\ref{eq-eloss}), with
$\bar{\om}^{1/2}\propto R^{-2/3}$. (As expected on physical grounds,
$\varepsilon(p_{T0},R)$ saturates at large $R$.)
We have checked that this particular law  is in good
numerical agreement with the numerical results in
Fig.~\ref{Fig:eloss-2}. Once again the scaling is almost perfect.

\subsection{Scaling violations and energy loss for full in-medium parton shower}
\label{sub:full-show}

As explained at length in section~\ref{sec:recall-static}
(see~\cite{Caucal:2019uvr} for additional details), the VLEs radiated
inside the medium act as new partonic sources for MIEs.
Each of these new sources will therefore contribute to the overall
energy lost by the jet (via radiation of MIEs at angles $\theta > R$).
This results in a significant increase of the total jet energy loss,
compared to that of a single parton evolving via MIEs only.
In this case, it is important to consider the energy lost by the jet
as a whole instead of the energy lost by just the leading parton.
In this section, we study how our results from
Ref.~\cite{Caucal:2019uvr} are modified by the 
expansion of the medium, notably in the context of the scaling
relation derived in section~\ref{sec:scaling-mies}.

As discussed in Sect.~\ref{sec:ps} (see Fig.~\ref{Fig:phase-space}),
in the presence of the longitudinal expansion, the energies and the
emissions angles of the VLEs occurring inside the medium
are constrained by Eq.~\eqref{omega0fin} and $\th\ge \th_c=2/\sqrt{\qhat(L)L^3}$.
Together with the medium-size boundary,
$ \om=\omega_L(\th)\equiv 2/(L\th^2)$, this defines the intersection
point at $\omega=\om_c$ and $\theta=\th_c$, cf.~\eqref{omegac}.
The corresponding boundaries and intersection point for the
``equivalent'' static medium are obtained by replacing $\gamma$ by $0$
and $\qhat(L)$ by $\qhat_{\rm stat}$, with $\qhat_{\rm stat}$ given by
Eq.~(\ref{qeff}). 

In practice, we define the total jet energy loss as difference between
the energy of the initial hard parton, $p_{T0}$, and the the energy of
the final reconstructed jet after evolving the hard parton including
both VLEs and MIEs.
To make sure this includes only the medium-induced energy loss, and
not also vacuum-like emissions outside the jet, we subtract the equivalent
average energy loss computed on vacuum jets.
Our results for the full-jet energy loss are shown by the solid lines
in Fig.~\ref{Fig:eloss}.
The main trend, for both static and
expanding media, is a steady growth of the jet energy loss with the
initial energy $p_{T0}$ and with the radius $R$, due to the increase
of the phase-space for in-medium VLEs with both
$p_{T0}$ and $R$.

Fig.~\ref{Fig:eloss} also shows a slight decrease of the energy loss
for an expanding medium ($\gamma=1$) compared to the equivalent static
one ($\gamma=0$).
This reduction becomes sizeable only for large
initial energy $p_{T0}\gtrsim 200$~GeV and/or very small values for
the jet radius $R\lesssim 0.2$.  A priori, these scaling violations may
have two sources: the $k_\perp$-broadening discussed in
section~\ref{sec:scaling-violations-broadening} and the different
phase-space available to VLEs inside the medium.
The smaller difference between the dotted lines than between the solid
lines in Fig.~\ref{Fig:eloss}
already suggests that the second effect --- the difference in the phase-space
for VLEs ---  dominates over the former.
We study this effect in more details below.

Strictly speaking, both the slope of the boundary in (\ref{omega0fin})
and the coordinates of the intersection point, differ between the two
scenarios.
However, within our leading logarithmic approximation for the VLEs,
only the ($\gamma$-dependent) change in the slope is under
control: multiplying $\omega_0(\th)$ and $\omega_L(\th)$ by an
arbitrary numerical prefactor of order one does not affect
the overall leading-logarithmic accuracy (recall the discussion
in the paragraph after \eqn{omega0}).
In particular, at double-logarithmic accuracy, we could have chosen
the prefactor in Eq.~(\ref{omega0fin}) such that the intersection
point be given by \eqref{omegac} with $\qhat(L)\to \qhat_{\rm stat}$
for both the expanding and the static media, or, conversely,
constructed the static medium with $\hat{q}=\hat{q}(L)$.

That said, the difference in slopes between the two scenarios
does matter at double-logarithmic accuracy.
Using Eq.~(\ref{omega0fin}), one can easily check that this difference
is such that the phase-space available for in-medium VLEs is
smaller for an expanding medium than for the ``equivalent''
static one (see Fig.~\ref{Fig:phase-space}).
Therefore, the number of VLEs inside the medium, and hence of sources
for MIEs, is smaller for an expanding medium, resulting in a smaller jet energy loss.

\begin{figure}
  \begin{subfigure}[t]{0.48\textwidth}
    \includegraphics[width=\textwidth,page=1]{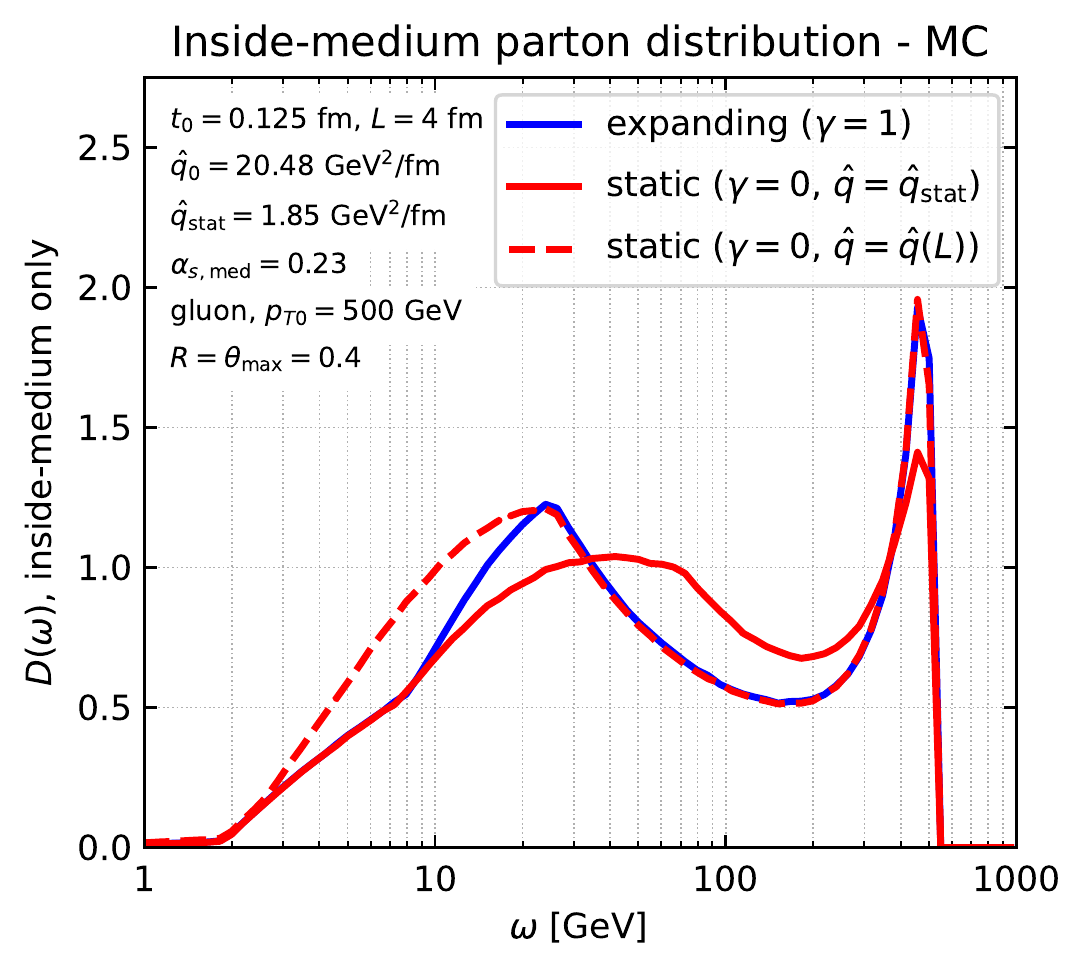}
    \caption{Monte-Carlo simulations}\label{fig:in-medium-ff-mc}
  \end{subfigure}
  \hfill
  \begin{subfigure}[t]{0.48\textwidth}    
    \includegraphics[width=\textwidth,page=2]{in-medium.pdf}
    \caption{Analytic (DLA) results}\label{fig:in-medium-ff-analytic}
  \end{subfigure}
  \caption{\small Parton energy distribution $D(\omega)
  \equiv { \omega(\rm d}N/{\rm d}\omega)$ in a parton shower 
  exclusively generated via VLEs inside the medium:
   (a) Monte Carlo simulations; (b) analytic results within the double-logarithmic
    approximation with fixed coupling.
    The leading parton is a gluon with $p_{T0}=500$~GeV
    and the jet radius is $R=0.4$.
    The solid blue lines correspond to the an expanding
    medium with $\gamma=1$; the red lines correspond to a static
    ($\gamma=0)$ medium, with either $\hat{q}=\hat{q}_\text{stat}$
    (solid lines), or $\hat{q}=\hat{q}(L)$ (dashed lines).
    The contribution from the leading parton (a 
    peak at $\omega\simeq p_{T0}=500$~GeV) is not included in
    the analytic plot.
  }\label{sec:in-medium-ff}
\end{figure}

To make this argument more concrete, we can directly look at the
energy ($\omega$) distribution of the partons produced only via
VLEs inside the medium (i.e.\ without MIEs and without the VLEs
outside the medium).
Fig.~\ref{sec:in-medium-ff} shows our results, for an initial parton
of 500~GeV. Fig.~\ref{fig:in-medium-ff-mc} is the result of our Monte
Carlo simulations and Fig.~\ref{fig:in-medium-ff-analytic} is the
result of the analytic calculation in the double-logarithmic
approximation (more details below).
Each plot includes three curves: the solid blue line corresponds to an
expanding medium, with $\gamma=1$ and our default medium parameters;
the two red curves correspond to two different choices for a static
medium, with different transport coefficients:\footnote{The
first choice, i.e. $\hat{q}=\hat{q}_\text{stat}$, is truly natural only in the
context of the MIEs, for which it guarantees the scaling property discussed
in Sect.~\ref{sec:scaling-mies}. In the present context of VLEs, both choices
look {\it a priori} reasonable.} $\hat{q}=\hat{q}_\text{stat}$,
cf. Eq.~(\ref{qeff}) (solid red line), and  $\hat{q}=\hat{q}(L)$
(dashed red line).
This last choice gives the same values of $\omega_c$ and $\theta_c$
as in the expanding medium.

In all cases, one sees a peak at large $\omega$ corresponding to the
leading parton, together with a continuous distribution extending
towards smaller $\omega$.
More importantly, we see a global increase in the number of sources in both
static media compared to the expanding one. This ultimately yields the
larger energy loss observed in the full MC simulations, as shown by the solid
lines in Fig.~\ref{Fig:eloss}.
This is more striking for the static medium where the values of
$\omega_c$ and $\theta_c$ have been chosen to agree with the expanding
medium, the red dashed line inf Fig.~\ref{sec:in-medium-ff}.
In this case, the static and expanding cases agree almost perfectly at
high $\omega$ ($\omega\gtrsim \omega_c = 25.6$~GeV). For
$\omega\lesssim \omega_c=25.6$~GeV,
the expanding medium yields a
distribution which is more suppressed than for the static one with
$\hat{q}=\hat{q}(L)$, due to the different slope of the phase-space boundary. 

We finally explain  our analytic results in
Fig.~\ref{fig:in-medium-ff-analytic}. We start from the
double-differential distribution in $\omega$ and $\theta$ computed in our
first paper~\cite{Caucal:2018dla},
\begin{equation}\label{T-pt-theta-dla}
  T(\omega,\theta)
  = \omega\theta^2 \,\frac{{\rm d}^2N}{{\rm d}\omega\,{\rm d}\theta^2}
  = \bar\alpha_s \textrm{I}_0\left(2\sqrt{\bar\alpha_s\ln\frac{p_{T0}}{\omega}\ln\frac{R^2}{\theta^2}}\right),
\end{equation}
with $\textrm{I}_0$ the modified Bessel function of rank 0.
This result is valid in the vacuum and in the fixed-coupling limit.
The inside-medium $\omega$ spectrum is obtained by
integrating~(\ref{T-pt-theta-dla}) over $\theta$ with the constraint that
the emission occurs inside the medium, i.e.\ satisfies both
Eq.~(\ref{omega0fin}) and $\theta>\theta_c$. The former constraint
dominates for $\omega<\omega_c$, while the latter  dominates
above $\omega_c$.
One finds
\begin{equation}\label{D-pt-dla}
  D(\omega)
  \equiv \omega \,\frac{{\rm d}N}{{\rm d}\omega}
  \simeq \sqrt{\bar\alpha_s \frac{L_\text{min}}{\ln\frac{p_{T0}}{\omega}}}\,
  \textrm{I}_1\left(2\sqrt{\bar\alpha_s\ln\frac{p_{T0}}{\omega}L_\text{min}}\right),
\end{equation}
with $\textrm{I}_1$ the modified Bessel function of rank 1 and
\begin{equation}
  L_\text{min} \equiv \begin{cases}
    \ln\frac{R^2}{\theta_c^2} & \text{ if }\omega>\omega_c,\\*[.2cm]
    \frac{1}{2-\gamma}\ln\frac{\omega^{3-\gamma}R^{4-2\gamma}}{2^{1-\gamma}\hat{q}_0
      t_0^\gamma} & \text{ if }\omega<\omega_c.
  \end{cases}
\end{equation}
This is the result plotted in Fig.~\ref{fig:in-medium-ff-analytic},
with an additional contribution $\delta(\omega-p_{T0})$ from the leading
parton omitted.
The analytic calculation captures, at least qualitatively, the
features seen in the Monte Carlo simulation, cf. Fig.~\ref{fig:in-medium-ff-mc}.

 \section{Jet quenching phenomenology in a longitudinally-expanding medium}
 \label{sec:pheno}
 
In this last section, we present MC simulations for three standard jet observables 
in ultra-relativistic heavy ion collisions: the nuclear modification factors for inclusive jet production,
the jet fragmentation function, and the $z_g$ and $\theta_g$
distribution obtained with the Soft Drop substructure tool.
Our aim is twofold.  On one hand, we would like to gauge the impact of
the medium expansion on the selected observables, by comparing it with
the corresponding predictions of the ``equivalent'' static-medium
scenario. On the other hand, we would like to demonstrate that adding
the longitudinal expansion to the general picture presented
in~\cite{Caucal:2018dla} still provides as good a phenomenological
description of these observables as that obtained for a static medium
in~\cite{Caucal:2019uvr,Caucal:2020xad}. As in these latter studies,
it is not our intention here to provide realistic fits of the
experimental data, but merely to show that overall physical picture
explains the salient features visible in these data.
 
 \subsection{The nuclear modification factor for jets $R_{AA}$} \label{sec:RAA}
 
 We start by computing the nuclear modification factor for the inclusive jet cross-section, $R_{AA}$, as a function of the jet transverse momentum $p_T$, following the ATLAS set-up~\cite{Aaboud:2018twu}. 
 
 Compared to our earlier work in~\cite{Caucal:2019uvr}, we have
 included in our simulation the effect of the nuclear PDF effects,
 which have been argued in~\cite{Pablos:2019ngg,Huss:2020dwe} to have
 a sizeable impact on $R_{AA}$ especially at large $p_T$.
 In practice, this is done by adding the EPPS (NLO) nuclear PDFs
 corrections~\cite{Eskola:2016oht} to the Born-level matrix elements
 used in our medium Monte Carlo medium simulations. Currently, these
 corrections are afflicted by large uncertainties, but in our MC
 calculations we have solely included their central
 values.\footnote{We also refer to Ref.~\cite{1832442} which studies
   the effect of nuclear PDFs and their uncertainties on $R_{AA}$,
   together with other sources of scale uncertainties.} Indeed, we
 have checked that the associated uncertainties could be absorbed into
 (slight) modifications of our free parameters, while still keeping
 them within physically reasonable ranges.
 
\begin{figure}[t] 
  \centering
  \begin{subfigure}[t]{0.48\textwidth}
    \includegraphics[width=\textwidth]{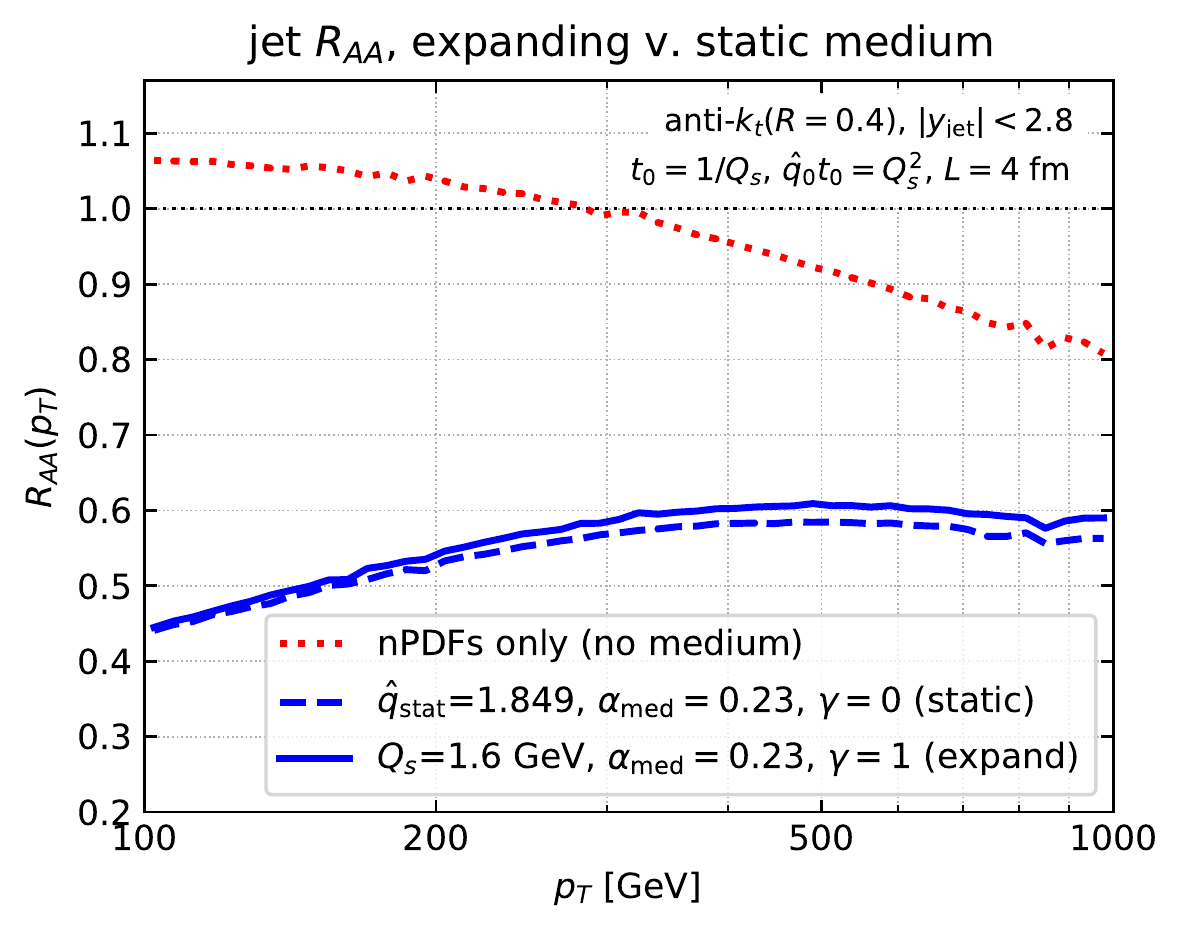}
    \caption{\small Effects of the nuclear PDFs alone (red dotted curve)
    and comparison between expanding and equivalent static medium (solid
    and dotted blue lines, respectively).}\label{Fig:RAA-pheno-1}
  \end{subfigure}
  \hfill
  \begin{subfigure}[t]{0.48\textwidth}
    \includegraphics[width=\textwidth]{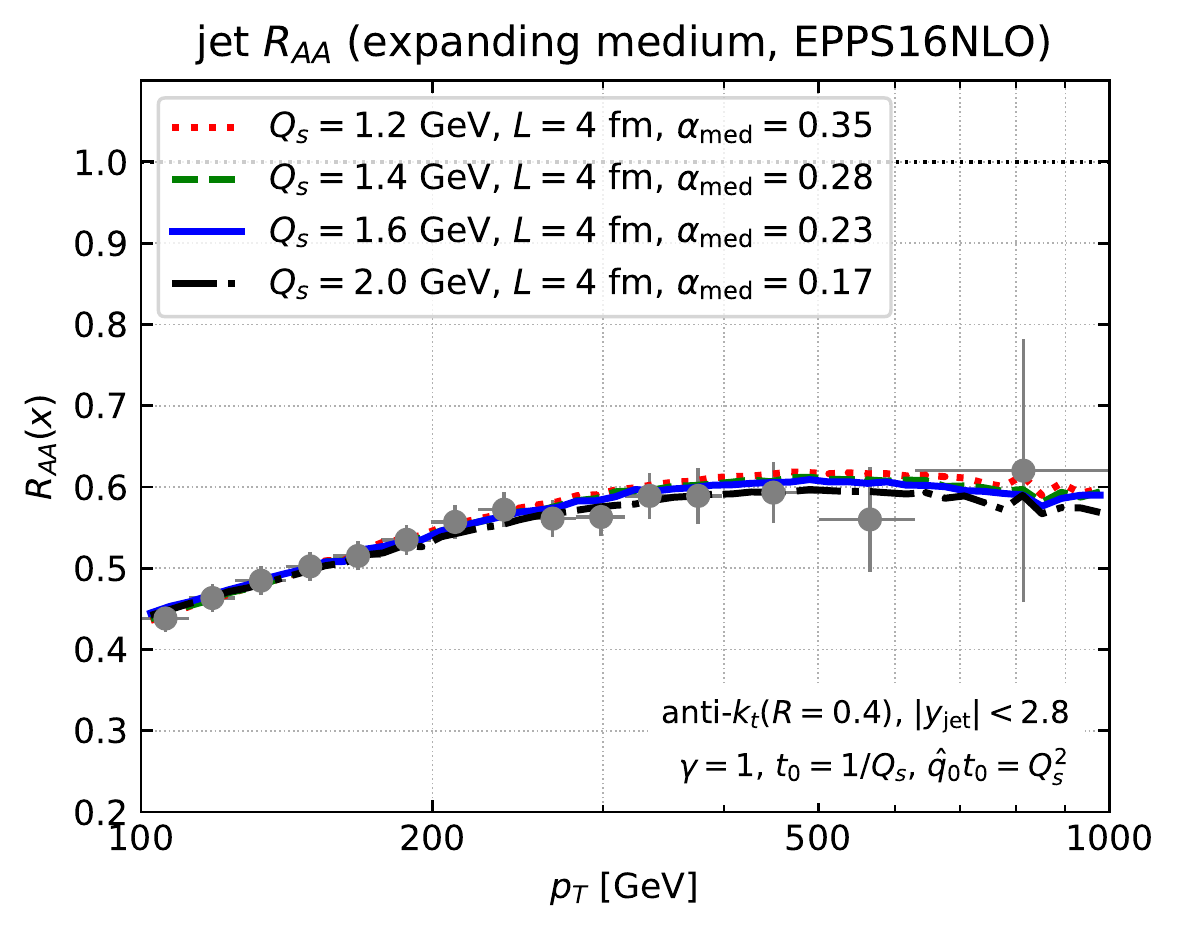}
    \caption{\small MC results for an expanding medium
      ($\gamma=1$) and nuclear PDFs vs.\ the ATLAS data~\cite{Aaboud:2018twu}. 
       The 4 sets of medium parameters
      displayed in this plot (cf.\ table~\ref{tab:parameters}) provide almost identical predictions.}\label{Fig:RAA-pheno-2}
  \end{subfigure}
  \caption{\small Monte-Carlo simulations for the nuclear modification factor $R_{AA}$ for
  inclusive jet production.}\label{Fig:RAA-pheno} 
\end{figure}

The red dotted curve in Fig.~\ref{Fig:RAA-pheno-1} shows the effect on $R_{AA}$ of the nuclear PDFs \textit{alone}, i.e.\ without any final-state quenching: the hard process is weighted by the nuclear PDFs,
 but the subsequent jet evolution occurs as in the vacuum.
 At large $p_T\gtrsim 500$ GeV, the nuclear PDFs tend to reduce the jet cross-section by $\sim 20\%$.
This is likely a consequence of  the EMC effect~\cite{Malace:2014uea}
i.e.\ of the suppression 
of the quark PDF at large  $x$ in nuclei compared to nucleons. Even if
this initial-state effect is sizeable, we remind the
reader that, in our picture, the crucial ingredient explaining the
flatness of $R_{AA}$ at high $p_T$ is the increase of the average
energy loss by the jets, due to the increase in the number of partonic
sources produced via VLEs inside the medium~\cite{Caucal:2019uvr}. This effect \textit{per se} (without nuclear
PDFs effects) was sufficient to provide a good description of the ATLAS data~\cite{Aaboud:2018twu} (within their uncertainty),
as shown in~\cite{Caucal:2019uvr}.
The inclusion of the nuclear PDF effects provides a further reduction
at large $p_T$, significantly improving the agreement with the ATLAS
data.

To isolate the effect of the medium expansion on $R_{AA}$, we show in Fig.~\ref{Fig:RAA-pheno-1}
the results of two MC simulations (both including the nuclear PDFs): one for a longitudinally-expanding 
medium with $\gamma=1$, $Q_s=1.6$ GeV, $L=4$~fm and $\alpha_{s,\rm
  med}=0.23$ (our default set of medium parameters, see the third line
in table~\ref{tab:parameters}) and the other one for the  ``equivalent'' static medium, with
$\qhat_{\rm stat}$ given by Eq.~(\ref{qeff}) and $L_{\rm stat}=L-t_0$.

The numerical results in Fig.~\ref{Fig:RAA-pheno-1} show only a mild
difference between the two scenarios, with $R_{AA}$ being slightly
larger for the expanding medium than for the static one and the
difference increasing slowly with the jet $p_T$.
Since the jet $R_{AA}$ is mainly controlled by the jet energy loss, 
this is in agreement with our previous discussion of
the scaling violations in Sect.~\ref{sec:scaling} (see
Fig.~\ref{Fig:eloss-1}) where we observed
a smaller average jet energy loss for the full in-medium parton shower
for an expanding medium compared to the equivalent static one, with a
stronger effect at large $p_T$.
The fact that
the net effect of the longitudinal expansion can be so well reproduced by an effective static-medium scenario
\textit{a posteriori} explains why it has been possible in~\cite{Caucal:2019uvr} to provide a rather 
good description of the ATLAS data for $R_{AA}$ within an oversimplified model assuming a static medium.

Let us now focus on more phenomenological studies of the $R_{AA}$
factor for a longitudinally-expanding medium.
Fig.~\ref{Fig:RAA-pheno-2} compares our Monte-Carlo predictions for
each of the four sets of medium parameters introduced in
table~\ref{tab:parameters} to the ATLAS measurement, showing an
excellent agreement.
For each set of parameter, the value of $\amed$ has been (manually)
adjusted to give a good description of the data.
For simplicity, we have only varied $Q_s$, setting $t_0=1/Q_s$ and
$\hat{q}_0t_0=Q_s^2$, keeping $\gamma=1$ and $L=4$~fm. These last two
parameters could have been varied as well.

The physical reason behind this degeneracy in our theoretical
description of $R_{AA}$ has been explained in detail in~\cite{Caucal:2019uvr}. In a nutshell, $R_{AA}$ is mainly sensitive to the energy loss via soft 
MIEs at large angles, i.e.\ to the branching scale $\ombr=\bar{\alpha}_{s,\rm med}^2\qhat_{\rm stat}L^2$.
However,
variations of the in-medium phase space for vacuum-like sources associated with variations of $\qhat_0$ and $t_0$ (through $Q_s$) can compensate the variations of $\ombr$ through $\alpha_{s,\rm med}$.
At this point, it is interesting to observe that the value of $\alpha_{s,\rm med}$ which is 
preferred by our phenomenological description of the $R_{AA}$ data is monotonously decreasing
with increasing $Q_s$, in qualitative agreement with the property of asymptotic freedom. (Indeed,
increasing $Q_s$ is tantamount to increasing the density of the medium, as obvious from the fact
that $\hat{q}_0=Q_s^3$.)

\subsection{Jet fragmentation function} \label{sec:FF}

We turn now to the discussion of the jet fragmentation function $\mathcal{D}(x)=\frac{1}{N_{\textrm{jets}}}\frac{\dif N}{\dif x}$, defined as the multiplicity of hadrons inside the jets per unit of longitudinal momentum fraction $x\equiv p_T\cos(\Delta R)/p_{T,\rm jet}$. Here, $p_T$ and  $\Delta R$ are respectively the transverse momentum and
the angle with respect to the jet axis of the measured hadron, while $p_{T,\rm jet}$ 
is the jet total transverse momentum. We denote by
$\mathcal{R}(x)\equiv \mathcal{D}^{\rm med}(x)/\mathcal{D}^{\rm
  vac}(x)$ the  associated nuclear modification factor.
The jet selection used in our MC analysis closely follows the experimental analysis by the ATLAS collaboration in~\cite{Aaboud:2018hpb}.

\begin{figure}[t] 
  \centering
  \begin{subfigure}[t]{0.48\textwidth}
    \includegraphics[width=\textwidth]{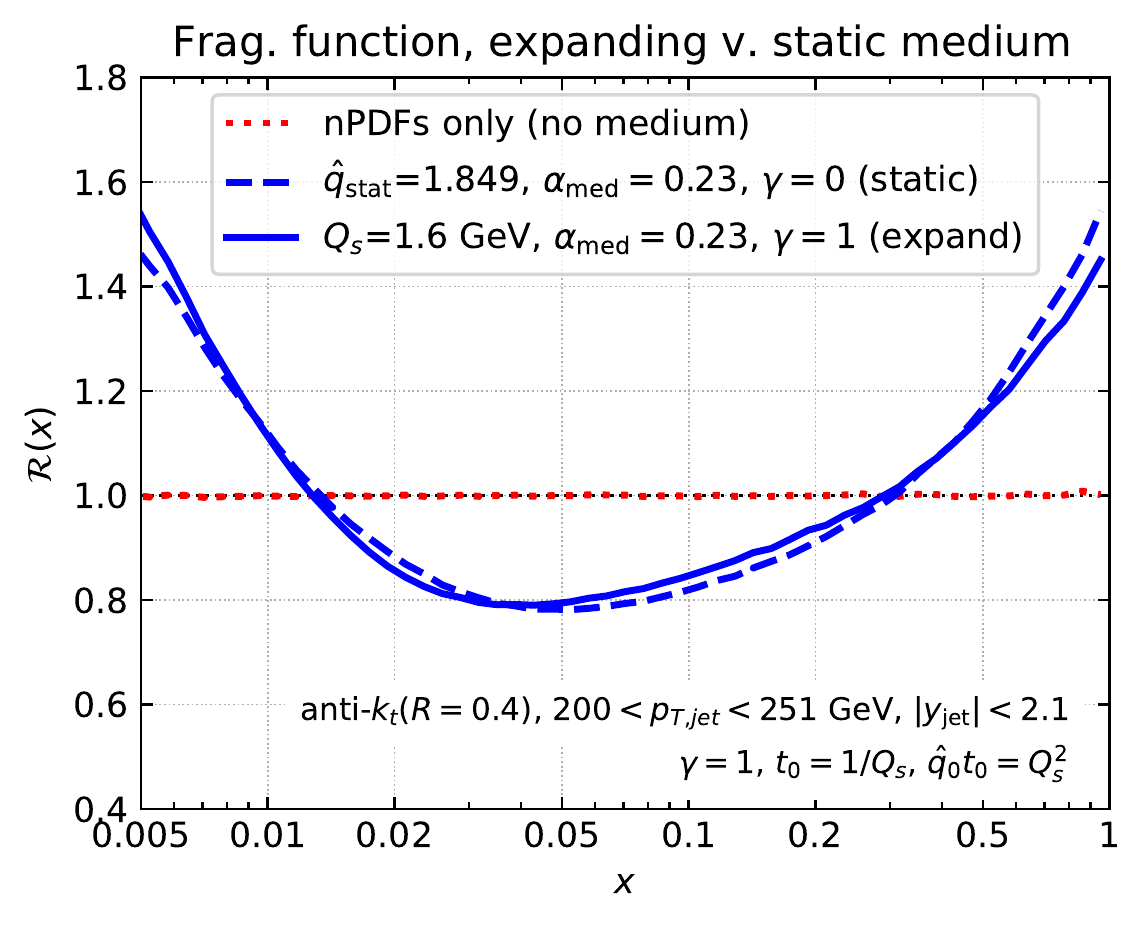}
    \caption{\small Effects of the nuclear PDFs alone (nearly-horizontal, red, dotted curve)
    and comparison between expanding and equivalent static medium (solid
    and dotted blue lines, respectively).}\label{Fig:ff-pheno-1}
  \end{subfigure}
  \hfill
  \begin{subfigure}[t]{0.48\textwidth}
    \includegraphics[width=\textwidth]{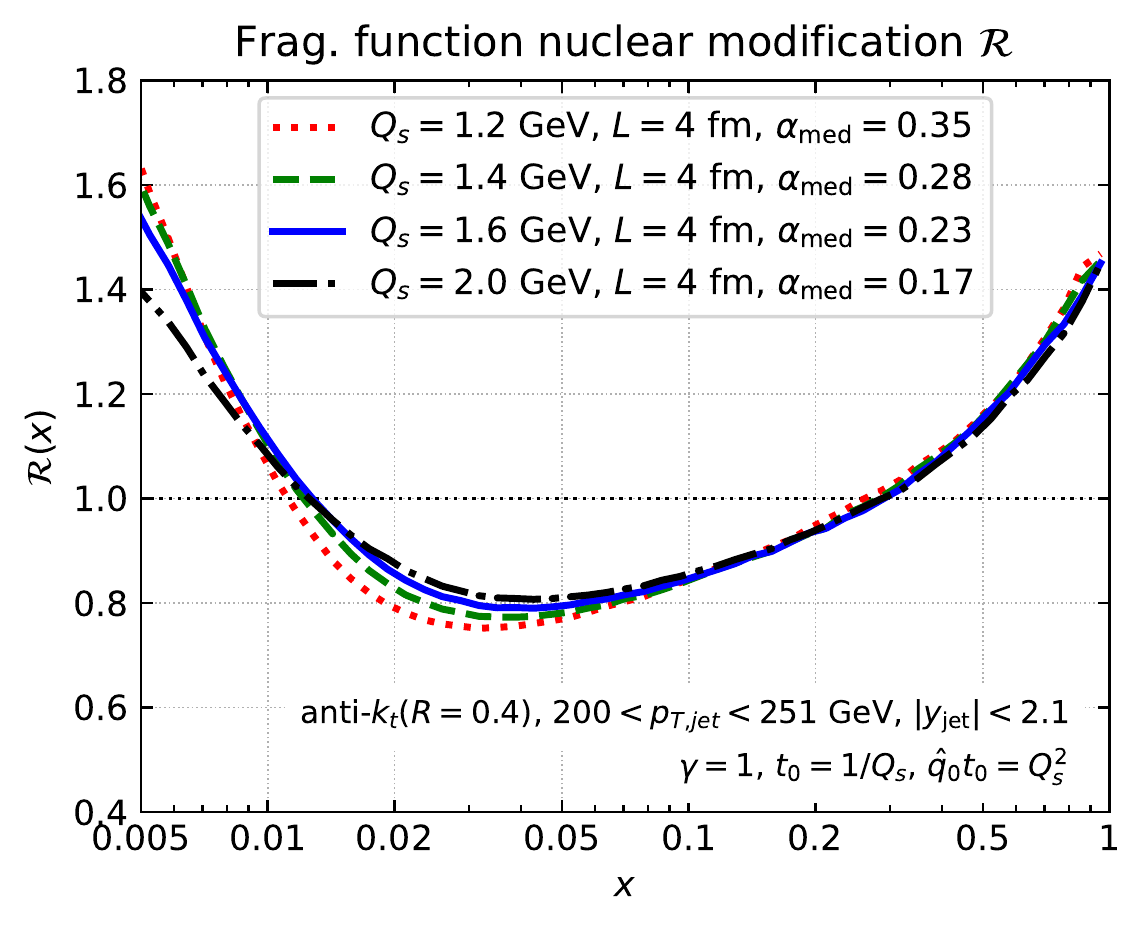}
    \caption{\small MC results for a Bjorken-expanding medium
      with $\gamma=1$ and for the 4 sets of medium parameters
      summarised in table~\ref{tab:parameters}.}\label{Fig:ff-pheno-2}
  \end{subfigure}
  \caption{\small Monte-Carlo calculation of the nuclear modification factor for the jet fragmentation function.}\label{Fig:ff-pheno} 
\end{figure}

In a previous paper~\cite{Caucal:2020xad}, we have studied the nuclear modification of the fragmentation function within our pQCD picture for the case of a static medium. To understand the effect of the longitudinal expansion on this observable, it is again enlightening to compare our MC results
for $\mathcal{R}(x)$ for the case of an expanding medium and for the ``equivalent'' static medium.
This comparison is shown in Fig.~\ref{Fig:ff-pheno-1}, for our default set of values for the free parameters 
(cf.\ the bold line in Table~\ref{tab:parameters}). 

The dotted red curve in this figure corresponds to a calculation which includes the nuclear PDFs in the initial state,
but no medium effects in the final state, in analogy with the dotted
red curve in Fig.~\ref{Fig:RAA-pheno-1}. Unlike in the case of $R_{AA}$, it appears that the nuclear PDFs have no effect
on the jet fragmentation function. This is likely related to the fact that the jet transverse momenta
involved in the present calculation of $\mathcal{R}(x)$ are relatively low,  $200 \le p_{T,\rm jet} \le 251$~GeV. 

Our MC results for $\mathcal{R}(x)$, including both nuclear PDFs and the in-medium effects,
are represented by the plain blue curve for the Bjorken-expanding
medium and by 
the dotted blue curve for the equivalent static medium. As for $R_{AA}$, the scaling looks nearly exact.
At large $x\gtrsim 0.2$, the fragmentation function enhancement
($\mathcal{R}(x)> 1$) has been shown in Ref.~\cite{Caucal:2020xad} to
be controlled by the energy loss by the jet together with the bias
introduced by the steeply falling initial spectrum which favours jets
losing less energy than average.
The almost-perfect scaling for the fragmentation at large $x$ therefore
stems from the equivalent almost-perfect scaling seen for $R_{AA}$ in
Fig.~\ref{Fig:RAA-pheno-1}.
This is a rather universal feature, that has been argued in model-independent
phenomenological studies~\cite{Spousta:2015fca} and is indeed verified in a variety of theoretical
descriptions, from weak to strong coupling~\cite{Milhano:2015mng,KunnawalkamElayavalli:2017hxo,Chesler:2015nqz,Rajagopal:2016uip,Casalderrey-Solana:2016jvj,Casalderrey-Solana:2018wrw,Casalderrey-Solana:2019ubu}.

The rather good scaling visible in Fig.~\ref{Fig:ff-pheno-1} at small $x\lesssim 0.02$ is likely to be fortuitous: within our effective theory at least,  it
is the result of the compensation between two scaling-violating effects, which act in opposite directions.
As argued in~\cite{Caucal:2020xad}, the small-$x$ part of the medium-modified fragmentation function is controlled by the multiplicity of in-medium VLEs and by the number of MIEs that remain inside the jet cone
after crossing the medium.
The latter effect tends to increase in an expanding medium compared to
the equivalent static one since the transverse momentum broadening
decreases (cf.\ Fig.~\ref{Fig:frag-func-2} and the discussion in
Sect.~\ref{sec:scaling-violations-broadening}).
The former effect decreases in the expanding medium since the associated phase-space is smaller
than for  the equivalent static medium (cf.\ Sect.~\ref{sub:full-show}).
The net effect visible in Fig.~\ref{Fig:ff-pheno-1} turns out to
be a mild increase. 

Turning to a more phenomenological analysis, we exhibit in Fig.~\ref{Fig:ff-pheno-2} our MC results for
the nuclear modification factor $\mathcal{R}(x)$ for jet
fragmentation, for the same four sets of medium parameters in
table~\ref{tab:parameters} that were already shown in Fig.~\ref{Fig:RAA-pheno-2} to offer a good description for the ATLAS data for $R_{AA}$. 
At large $x\gtrsim 0.2$, the four curves are nearly overlapping with each other, a property associated 
in~\cite{Caucal:2020xad} with the strong correlation between hard-fragmenting jets and $R_{AA}$.
At the small-$x$ end of the spectrum, the dispersion between the
different curves is more pronounced albeit still small. This
reflects the complexity of the physical mechanisms at work in that
regime.

In view of the theoretical uncertainties inherent to our current framework,
which are especially important for the fragmentation function at small $x$
(see again~\cite{Caucal:2020xad}),
we do not show an  explicit comparison between our results and the 
LHC data for nuclear effects on jet fragmentation. That said,
it  is reassuring to observe that all our curves in Fig.~\ref{Fig:RAA-pheno-2} show the
same qualitative features as the respective data   (see e.g.~\cite{Aaboud:2018hpb}), 
that is, a pronounced nuclear enhancement at both small $x$ and large
$x$, together with a nuclear suppression at intermediate values of $x$.

\subsection{Jet substructure observables} \label{sec:substructure}

To conclude this survey of jet quenching observables in a
longitudinally expanding medium, we study the Soft
Drop $z_g$ and $\th_g$
distributions~\cite{Larkoski:2014wba,Larkoski:2015lea} for $\beta=0$,
and their respective nuclear modification factor $\mathcal{R}(z_g)$,
$\mathcal{R}(\th_g)$. Being infrared and collinear safe ($\theta_g$) or
at least Sudakov-safe ($z_g$, see Ref.~\cite{Larkoski:2015lea}), these
observables are expected to be better-controlled in perturbation
theory than the fragmentation function. In particular, they
are less sensitive to non-perturbative hadronisation corrections which
are not included in our Monte Carlo.
For brevity, we only show results for the longitudinally-expanding
medium with $\gamma=1$.
We have however checked that the scaling between an expanding medium
and the equivalent static one works well for these substructure
observables, 

 \begin{figure}[t] 
  \centering
  \begin{subfigure}[t]{0.48\textwidth}
    \includegraphics[width=\textwidth]{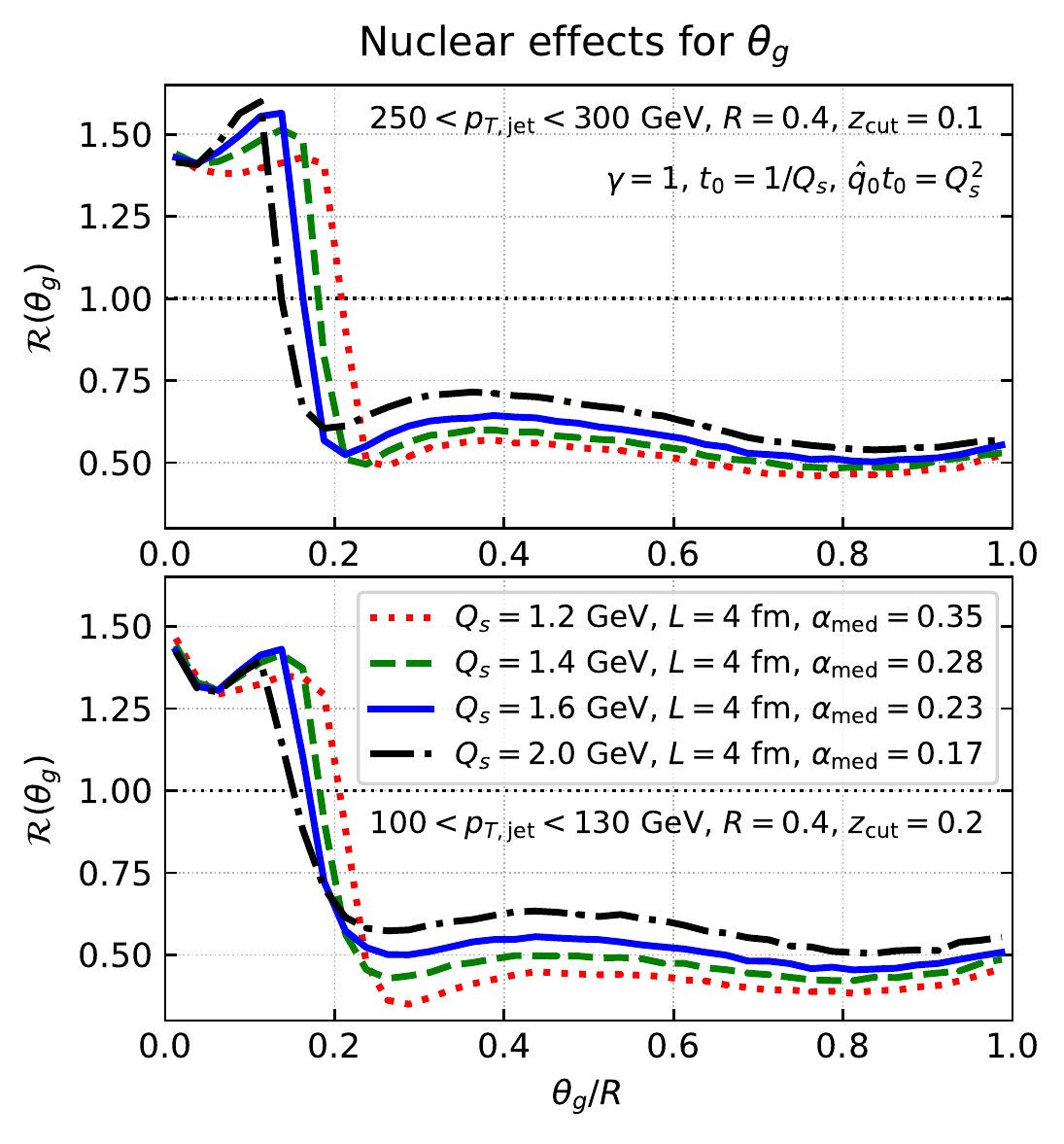}
    \caption{\small $\theta_g$ distributions.}\label{Fig:grooming-pheno-1}
  \end{subfigure}
  \hfill
  \begin{subfigure}[t]{0.48\textwidth}
    \includegraphics[width=\textwidth]{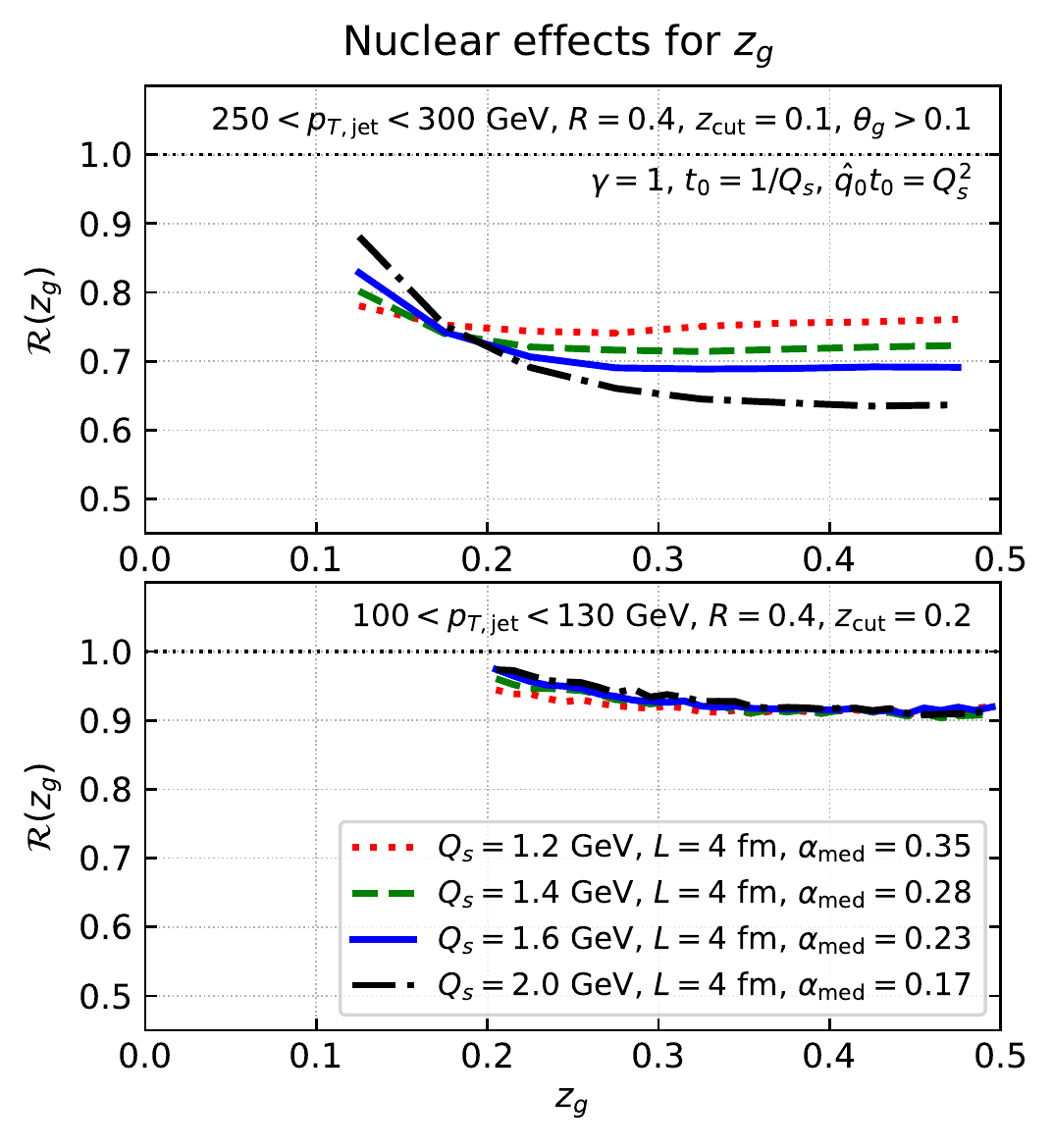}
    \caption{\small $z_g$ distributions.}\label{Fig:grooming-pheno-2}
  \end{subfigure}
  \caption{\small Monte-Carlo simulations of the nuclear modification
    factor for distributions after Soft Drop: (a) the groomed jet
    radius $\theta_g$ and (b) the momentum fraction $z_g$. 
  The Soft Drop parameter $\beta$ is set to $\beta=0$.  
  For both distributions, the upper panel corresponds
  to $z_{\rm cut}=0.1$ and $250 <p_{T,\rm jet}<300$~GeV, whereas the lower panel has
 $z_{\rm cut}=0.2$ and $100 <p_{T,\rm jet}<130$~GeV.
 As in the ALICE~\cite{Acharya:2019djg} and CMS measurements~\cite{Sirunyan:2017bsd}, an additional angular cut-off $\theta_g>0.1$ is imposed for the $z_g$ distribution calculated with $z_{\rm cut}=0.1$. We use
 the same sets of values for the medium parameters
  as in Figs.~\ref{Fig:RAA-pheno-2}--\ref{Fig:ff-pheno-2} (cf. table~\ref{tab:parameters}).}\label{Fig:grooming-pheno} 
\end{figure}

Let us first consider the groomed $\th_g$ distribution shown in Fig.~\ref{Fig:grooming-pheno-1} for two choices for the jet selection in $p_{T,\rm jet}$ and for the Soft Drop parameter $\zcut$: 
$250 <p_{T,\rm jet}<300$~GeV and $\zcut = 0.1$ (upper panel), and, respectively, $100
<p_{T,\rm jet}<130$~GeV and $\zcut=0.2$ (lower panel). 
In both cases, one observes a suppression of large--$\th_g$ jets and an enhancement of small--$\th_g$ jets,
with the transition between the two types of behaviour occurring
slightly below $\th_g=0.1$ (i.e.\ for $\theta_g/R\sim 0.2$).

This behaviour indicates that the opening angle distribution for jets
emerging from the plasma within any specified range of energies has
been pushed toward smaller angles, comparing to $pp$ jets with the
same energies.
This trend is a rather generic feature, which has been observed in a variety of theoretical descriptions, at both weak~\cite{Ringer:2019rfk} and strong coupling~\cite{Chesler:2015nqz,Rajagopal:2016uip}, 
and also in hybrid models~\cite{Casalderrey-Solana:2014bpa,Hulcher:2017cpt,Casalderrey-Solana:2019ubu}. The
ultimate reason for this narrowing of jets in the medium is that
small-angle jets suffer less energy loss and jets with a higher
initial energy are less frequent (due to the steeply falling initial spectrum).

To understand our results in Fig.~\ref{Fig:grooming-pheno-1} in more detail and,
in particular, the peculiar transition occurring around $\th_g=0.1$, it is important to elucidate the precise
mechanisms relating the (sub)jet angular opening to energy loss in our
picture. There is first a rather
generic mechanism, present in all pQCD-based approaches:  the medium introduces a
bias towards quark-initiated jets,\footnote{This bias is also responsible for most of the nuclear
enhancement seen in the jet fragmentation function at large $x$~\cite{Spousta:2015fca,Milhano:2015mng,KunnawalkamElayavalli:2017hxo,Caucal:2020xad}, 
cf.\ Fig.~\ref{Fig:ff-pheno-2}.} which lose less energy than the gluon jets,
thus leading to a narrowing of the angular distribution~\cite{Ringer:2019rfk},
since quark jets are ``narrower'' than gluon jets.\footnote{A simple
  leading-log, fixed-coupling, estimation of the average 
groomed $\th_g/R$ of quark or gluon initiated jets gives, for $\beta=0$, $\langle
\th_g/R\rangle\simeq L_c/(1+L_c)$ with
$L_c=2\frac{\alpha_sC_R}{\pi}\log(1/\zcut)$~\cite{Larkoski:2014wba}, which
increases as $C_R$ increases.}
Conversely, the transverse broadening of the two subjets selected by Soft Drop leads 
to a widening of the $\th_g$ distribution~\cite{Ringer:2019rfk}. 
This effect too is present in our MC calculations but turns out to be
numerically small.

The dominant mechanism behind the sharp transition observed
in the $\th_g$--distribution in Fig.~\ref{Fig:grooming-pheno-1} is
colour decoherence and the associated angle $\theta_c$,
cf.\ Eq.~(\ref{omegac}).
The physical picture is as follows: when the two subjets selected by
Soft Drop have an angular separation $\theta_g> \theta_c$, each subjet
loses energy {\it independently}, whereas when $\th_g<\theta_c$ the
energy is lost by the {\it parent} subjet.
Since the jet energy loss increases with the number of partonic
sources for MIEs, jets with $\th_g>\th_c$ lose more energy than jets
with $\th_g<\th_c$.
This explains the suppression of the $\theta_g$ distribution for
$\th_g> \th_c$ and the enhancement for $\th_g< \th_c$.
This shows that the medium acting as a filter towards
coherent jets, by reducing the number of two-prongs jets with
$\th_g\gg\th_c$.
A similar mechanism is also present in strong coupling models, where
the role of the coherence angle is played by the plasma resolution
length~\cite{Casalderrey-Solana:2014bpa,Hulcher:2017cpt,Casalderrey-Solana:2019ubu}.
Numerically, the transition in Fig.~\ref{Fig:grooming-pheno-1} is
indeed seen at $\theta_g\simeq \theta_c$, with $\theta_c$ varying between $0.08$ and $0.05$ 
when $Q_s$ varies between $1.2$ GeV and $2$ GeV, cf.\
table~\ref{tab:parameters}, including the expected shift of the
transition point towards smaller $\theta_g$ when $\theta_c$ decreases
($Q_s$ increases).

On the experimental side, the narrowing of jets in the medium has been measured by the ALICE collaboration~\cite{Mulligan:2020cnp}. The results are in qualitative agreement with our theoretical predictions shown Fig.~\ref{Fig:grooming-pheno}, as a suppression of large--$\th_g$ jets and an enhancement of small--$\th_g$ jets are clearly visible in spite of the large statistical uncertainties. The transition in the  data
occurs around $\th_g/R\approx0.2\div0.3$ and it does not look as sharp
as in our MC results in  Fig.~\ref{Fig:grooming-pheno-1}.
This should be expected, given that the experimental results represent
an average over various medium geometries (hence, over various values
for $\th_c$).  Besides, also on the theory side, we expect that the
sharpness of the transition will be smoothed out by subleading
perturbative contributions and additional effects like hadronisation,
not yet included in our simulations.
The above discussion however suggests that a more precise measurement of the $\th_g$ distribution at
small $\th_g\ll R$  could give us a direct experimental access to the plasma coherence angle $\th_c$,
at least in an average sense.

Finally, the groomed $z_g$--distribution in a Bjorken-expanding medium is presented Fig.~\ref{Fig:grooming-pheno-2} for the same two choices of jet selection and Soft Drop parameters as for the 
$\th_g$--distributions in Fig.~\ref{Fig:grooming-pheno-1}. For the
upper panel we have imposed the additional restriction that the
groomed angle should satisfy $\th_g>\th_{\rm cut}=0.1$, as is common
in several experimental analyses.
Within our pQCD picture, the physical content of the $z_g$
distribution has been explored in~\cite{Caucal:2019uvr} for the case
of a static medium. The medium expansion does not change the physical
mechanisms at the origin of the nuclear modifications.  The overall
suppression of the distribution, i.e.\ the fact that
$\mathcal{R}(z_g)<1$ for all values of $z_g$, is again a consequence
of the in-medium suppression of jets having a large-angle hard
substructure. This suppression is therefore less pronounced for
inclusive $\th_g$ jets (lower panel) than for jets with $\th_g>0.1$ (upper
panel). On the other hand, the increase of $\mathcal{R}(z_g)$ when
decreasing $z_g$, especially below $z_g\sim 0.2$, is due to relatively
hard MIEs triggering the Soft Drop condition. As visible in the upper
panel, this increase becomes more pronounced as $Q_s$ increases (for
$L$ fixed). This is so since $\th_c=2/(Q_sL)$ decreases with
increasing $Q_s$ (see also table~\ref{tab:parameters}), meaning that
the relatively hard emissions with $\th\sim\th_c$ are more likely to
remain inside the jet cone.
Additionally, $\omega_c=\frac{1}{2}Q_s^2L$ increases with increasing
$Q_s$, making MIEs more likely to pass the Soft Drop $z_\text{cut}$ condition.
  
\section{Conclusion}
\label{sec:conc}

In this paper, we have extended our pQCD approach 
to the evolution of a jet in a dense quark-gluon plasma,
as developed in Refs.~\cite{Caucal:2018dla,Caucal:2019uvr,Caucal:2020xad},
to the case of a medium which undergoes longitudinal expansion.
We have demonstrated that the factorisation in time between vacuum-like emissions and medium-induced emissions remains valid in the expanding plasma
and established the new phase space for in-medium VLEs
to leading logarithmic accuracy. Regarding the medium-induced radiation, we have focused on
the relatively soft emissions with energies $\omega\ll\omega_c$, 
for which the emission process can be approximately treated as local in time. 
This locality allowed us to absorb the medium expansion into an effective emission rate which
involves the value of the jet quenching parameter at the time of the emission.
Our Monte Carlo parton shower, initially designed for a static plasma~\cite{Caucal:2019uvr}, 
has been extended to take into account the three main modifications introduced by the longitudinal expansion, 
namely, the modified phase space for VLEs, the change in the rate for transverse momentum broadening,
and that in the emission rate for MIEs.

From a conceptual point of view, we have made the elementary, yet important, observation,
that the locality of soft MIEs leads to exact scaling properties between expanding and static media 
for the parton energy distributions (integrated over the emission angles) in
medium-induced cascades. This has allowed us to define an ``equivalent
static medium'' for a given expanding scenario, with the equivalence being strictly true for MIEs only.
We have used this ``equivalent'' static medium as a benchmark for ``apple-to-apple'' comparisons
between the jet properties (energy loss, intra-jet multiplicities, jet substructure) for an expanding and
a static medium, respectively.
Our key result is that the scaling property is only mildly violated by processes which have a different
scaling with $\hat q(t)$,  such as the transverse momentum broadening via multiple soft scattering,
or the phase-space for VLEs occurring inside the medium (which act as sources for medium-induced
radiation).

On the phenomenological side, we have presented new Monte Carlo simulations for the case
of an expanding plasma, which cover the nuclear
modification factors for inclusive jet production (the jet $R_{AA}$), the jet fragmentation function, and
the Soft Drop distributions in the groomed radius $\th_g$ and the momentum sharing fraction $z_g$. The good qualitative agreement that we previously found between these jet observables computed in a static medium and the LHC data turns out to remain in place after including the medium expansion. This is so because
of the mildness of the scaling violations w.r.t.\ the ``equivalent'' static medium. This consolidates the pQCD foundations of our picture for jet evolution, since it confirms that the salient features of the nuclear
effects on all the observables that we have investigated
are driven by perturbative effects and are only slightly sensitive to the details of the bulk evolution.
Our studies of the jet $R_{AA}$ modification factor also confirm that
the inclusion of nuclear PDF effects improve the agreement with the
ATLAS data at large $p_{T,\text{jet}}$.

That said, there is still a large room for improvement in our current
implementation of an in-medium parton shower. As emphasised in several places, 
both the collisions inside the medium and the medium-induced radiations are considered as
relatively soft. Even though preliminary studies seem to indicate a
minor effect on the jet $R_{AA}$, a proper treatment of the single hard scattering regime~\cite{Gyulassy:2000er}
and of the medium-induced spectrum at larger energies  $\om\gtrsim \om_c$
(perhaps along the lines of the recent studies in~\cite{Mehtar-Tani:2019tvy,Mehtar-Tani:2019ygg,Barata:2020rdn})
is clearly needed and is a part of our plans for the future.
At this point, it is interesting to notice that the scaling relation put forward in~\cite{Baier:1998yf,Salgado:2002cd,Salgado:2003gb,Adhya:2019qse} for the relatively
hard ($\om\gtrsim \om_c$) medium-induced emissions differs from ``our" scaling by merely a
factor of $2$ (in the value of the transport coefficient for the ``equivalent''  medium), 
thus suggesting that the scaling violations may remain small
even after including this contribution.

We have intentionally focused on jet observables for which the medium geometry can, to a large extent, be absorbed into an effective jet path length $L$. Incorporating a realistic collision geometry is left for future work, as well as a systematic study of jet observables sensitive to it, such as the centrality dependence of $R_{AA}$, or the dijet asymmetry. Finally, even if perturbative mechanisms seem to drive the medium modifications of jet properties, it is known that bulk-related observables such as the medium response to the jet propagation, or the transverse expansion of the quark-gluon plasma, have a sizeable impact in the soft sector of some observables, notably on the fragmentation function and on the jet shapes~\cite{Casalderrey-Solana:2016jvj,Tachibana:2017syd,KunnawalkamElayavalli:2017hxo,Chen:2017zte,Tachibana:2020mtb}. Adding all such effects goes far beyond the current implementation of our Monte-Carlo, yet this is clearly needed in order to develop a
realistic event generator for the study of jets in heavy-ion collisions. 
This discussion can be viewed as our road map for the next years.

\section*{Acknowledgements} 

We are grateful to Al Mueller for insightful discussions which have influenced the critical
early stages of this paper.
The work of P.C., E.I.\ and G.S.\ is supported in part by the Agence
Nationale de la Recherche project ANR-16-CE31-0019-01.
G.S.\ is also partially supported by the Agence
Nationale de la Recherche project ANR-15-CE31-0016.

\appendix

\section{Angular structure of medium-induced cascades in expanding media}
\label{app:A}

In this Appendix, we shall demonstrate the results asserted in Sect.~\ref{sec:scaling} regarding the violation of exact scaling between the expanding scenario and the equivalent static medium, due to transverse momentum broadening in medium-induced cascades.  Our subsequent treatment of multiple
medium-induced branchings largely follows the corresponding discussion in
Refs.~\cite{Blaizot:2013hx,Blaizot:2013vha,Blaizot:2014ula,Blaizot:2014rla},
 that we shall extend to the case of an expanding medium by simply replacing $\hat q\to \hat q(t)$,
both in the emission rate and  in the rate for diffusion in transverse momentum.


\subsection{Exact scaling for $k_\perp$-inclusive parton distributions}

Let us first explain in more mathematical terms what we mean by this exact scaling. As emphasised in Sect.~\ref{sec:scaling}, this scaling property only refers to the parton distribution produced by medium-induced cascades which are inclusive w.r.t.\ the transverse momentum of emission. The evolution equation for the $k_\perp$-inclusive parton distribution at time $t$, $D(x,t)\equiv x \dif N/\dif x$ is \cite{Blaizot:2013hx}
\begin{equation}\label{D-evol}
 \frac{\partial D(x,t)}{\partial t}=\bar{\alpha}_s\sqrt{\frac{\qhat(t)}{E}}\int\dif z\,\mathcal{K}(z)\left[\sqrt{\frac{z}{x}}D\left(\frac{x}{z},t\right)-\frac{z}{\sqrt{x}}D(x,t)\right],\qquad \mathcal{K}(z)=\frac{(1-z+z^2)^{5/2}}{(z(1-z))^{3/2}}.
\end{equation}
Without loss of generalities, this equation is written for purely
gluonic cascades.
The change of variable
\begin{equation}\label{eq:tau-scaling-transformation}
 \tau = \bar{\alpha}_s\int_{t_0}^t\dif t'\,\sqrt{\frac{\qhat(t')}{E}}
\end{equation}
enables one to rewrite~\eqref{D-evol} in terms of the dimensionless
quantities $\tau$ and $x$:
\begin{equation}\label{D-evol-tau}
  \frac{\partial \widetilde{D}(x,\tau)}{\partial \tau}=\int\dif z\,\mathcal{K}(z)\left[\sqrt{\frac{z}{x}}\widetilde{D}\left(\frac{x}{z},\tau\right)-\frac{z}{\sqrt{x}}\widetilde{D}(x,\tau)\right].
\end{equation}
Consequently, the solutions of~\eqref{D-evol} are of the form 
$D(x,t)=\widetilde{D}(x,\tau(t))$,
with $\widetilde{D}(x,\tau)$ a solution of~\eqref{D-evol-tau}.\footnote{Although this discussion implicitly
  assumes that $E<\tilde{\om}_c$, the scaling
  transformation~\eqref{eq:tau-scaling-transformation} trivially holds
  also in the regime $E>\tilde{\om}_c$.}
If one fixes the initial time $t_0$ and final time $L$ of the
evolution, the $k_\perp$-inclusive parton distribution in an expanding
medium is the same as in a static one provided one has
\begin{equation}\label{tau-eq}
  \tau_{\rm exp}(L)=\tau_{\rm stat}(L)
  \qquad \Longleftrightarrow \qquad
  \int_{t_0}^L\dif t\,\sqrt{\qhat(t)}=\sqrt{\qhat_{\rm stat}}(L-t_0)
\end{equation}
giving the condition \eqref{qeff}. We refer to
Eq.~\eqref{tau-eq} as the scaling relation between expanding and
static media with $\qhat_{\rm stat}$ the quenching parameter associated with the equivalent static medium.

For future analytic calculations, it is helpful to note that when the branching kernel $\mathcal{K}(z)$ is approximated by $\mathcal{K}_0(z)=(z(1-z))^{3/2}$, there exists an analytic solution of \eqref{D-evol-tau} given by:
\begin{equation}\label{sol-simple}
 \widetilde{D}(x,\tau)\equiv\frac{\tau}{\sqrt{x}(1-x)^{3/2}}\exp\left(\frac{-\pi\tau^2}{1-x}\right).
\end{equation}

\subsection{Transverse momentum dependence of parton distributions}

To calculate observables sensitive to the angular distribution of the
medium-induced cascade (such as the fragmentation function for a jet
of radius $R$), one needs the fully differential distribution 
\begin{equation}\label{kt-distribution}
 D(x,k_\perp,t)\equiv(2\pi)^2x\frac{\dif N}{\dif x \dif^2 k_\perp}.
 \end{equation}
The evolution equation for this quantity is given by a generalisation of~\eqref{D-evol} with a new term accounting for the transverse diffusion of emissions due to momentum broadening \cite{Blaizot:2013vha}:
\begin{equation}\label{D-evol-kt}
 \frac{\partial D(x,k_\perp,t)}{\partial t}=\frac{\qhat(t)}{4}\nabla_\perp^2D(x,k_\perp,t)+\bar{\alpha}_s\sqrt{\frac{\qhat(t)}{E}}\int\dif z\mathcal{K}(z)\left[\frac{1}{z^2}\sqrt{\frac{z}{x}}D\left(\frac{x}{z},k_\perp,t\right)-\frac{z}{\sqrt{x}}D(x,k_\perp,t)\right].
\end{equation}
Because of the diffusion term, it is clear that the change of variable~\eqref{reduced-time-exp} does not cancel all the $\qhat(t)$ dependence in Eq.~\eqref{D-evol-kt}.

Instead of studying the full $k_\perp$ distribution, let us study the
average transverse momentum of gluons at the end of the evolution,
defined as
\begin{equation}\label{average-kt-def}
  \bar{k}_\perp^{\,2}(x)
  = \frac{\int \dif^2 k_\perp\,k_\perp^2D(x,k_\perp,L)}{\int \dif^2 k_\perp\,D(x,k_\perp,L)}
  \equiv \frac{H(x,L)}{D(x,L)}
  \equiv \frac{x^2 W(x,L)}{D(x,L)}.
\end{equation}
Following~\cite{Blaizot:2014ula}, we have defined $H(x,t)=\int
\dif^2k_\perp k_\perp^2 D(x,k_\perp,t)$ to be the first moment of the distribution \eqref{kt-distribution} and $W(x,t)=H(x,t)/x^2$. From~\eqref{D-evol-kt}, one easily gets the following evolution equation for $W(x,t)$
\cite{Blaizot:2014ula}:
\begin{equation}
 \frac{\partial W(x,t)}{\partial t}=\frac{\qhat(t)}{x^2}D(x,t)+\bar{\alpha}_s\sqrt{\frac{\qhat(t)}{E}}\int\dif z\,\mathcal{K}(z)\left[\sqrt{\frac{z}{x}}W\left(\frac{x}{z},t\right)-\frac{z}{\sqrt{x}}W(x,t)\right]
\end{equation}
After the change of variable \eqref{reduced-time-exp}, this equation reduces to
 \begin{equation}\label{W-eq-tau}
 \frac{\partial \widetilde{W}(x,\tau)}{\partial \tau}=\frac{\sqrt{E\qhat(t(\tau))}}{\abar x^2}\widetilde{D}(x,\tau)+\int\dif z\,\mathcal{K}(z)\left[\sqrt{\frac{z}{x}}\widetilde{W}\left(\frac{x}{z},\tau\right)-\frac{z}{\sqrt{x}}\widetilde{W}(x,\tau)\right],
\end{equation}
where the remaining $\qhat$ dependence is a consequence of the scaling
violation caused by transverse momentum broadening. Except for the
diffusion term, this equation is the same as Eq.~\eqref{D-evol-tau},
meaning that one can find a solution to \eqref{W-eq-tau} if a solution
of Eq.~\eqref{D-evol-tau} is known, using convolution methods. Namely,
if $\widetilde{D}(x,\tau)$ is a solution of~(\ref{D-evol-tau}) and
$\tau_{\rm max}\equiv \tau(L)$, we have 
\begin{equation}\label{W-sol}
 \widetilde{W}(x,\tau_{\rm max})=\frac{\sqrt{E}}{\abar}\int_0^{\tau_{\rm max}}\dif \tau\int_x^1\frac{\dif y}{y}\,\widetilde{D}\left(\frac{x}{y},\frac{\tau_{\rm max}-\tau}{\sqrt{y}}\right)\frac{\sqrt{\qhat(t(\tau))}}{y^2}\widetilde{D}(y,\tau),
\end{equation}
which generalises the corresponding result for a static medium \cite{Blaizot:2014ula}.
For the simplified kernel $\mathcal{K}_0(z)$, the function $\widetilde{D}$
is given by Eq.~\eqref{sol-simple} so that one can numerically
evaluate the double integral \eqref{W-sol}. The result for
$\bar{k}_\perp^2(x)$ is shown Fig.~\ref{Fig:frag-func-2}.

\subsection{Average transverse momentum in the multiple-branching
  regime.} 

It is enlightening to get the behaviour of the exact solution
\eqref{W-sol} at small $x\ll \abar^2\tilde{\om}_c$, i.e.\ in the
multiple branching regime, assuming the simplified kernel
$\mathcal{K}_0(z)$.
In the small $x$ limit, the $y$-integration is dominated by small values of $y$, $1\gg y\gtrsim x$, so that the exponential in the function $\widetilde{D}(x/y,(\tau_{\rm max}-\tau)/\sqrt{y})$ fixes $\tau\simeq \tau_{\rm max}$ in the $\tau$ integral~\cite{Blaizot:2014ula}. Using $\widetilde{D}(y,\tau_{\rm max})/\widetilde{D}(x,\tau_{\rm max})\simeq \sqrt{x/y}$ at small $x$ and $y$, one gets
\begin{equation}
 \bar{k}_\perp^2(x)\simeq x^2\frac{\sqrt{E}}{\abar}\int_0^{\tau_{\rm max}}\dif \tau\int_x^1\frac{\dif y}{y}\,\widetilde{D}\left(\frac{x}{y},\frac{\tau_{\rm max}-\tau}{\sqrt{y}}\right)\frac{\sqrt{\qhat(t(\tau_{\rm max}))}}{y^2}\sqrt{\frac{x}{y}}.
\end{equation}
Since $t(\tau_{\rm max})=L$, one notices that $\qhat$ is automatically
set to its final value $\qhat(L)$.
Changing variables to $u=x/y$ and then  $\tau'=(\tau_{\rm max}-\tau)\sqrt{u/x}$, one easily gets
\begin{equation}\label{eq:avg-kperp-after-changes}
 \bar{k}_\perp^2(x)\simeq \frac{\sqrt{\qhat(L) xE}}{\abar}\int_x^1\dif u\,u\int_0^{\tau_{\rm max}\sqrt{\frac{u}{x}}}\dif \tau'\,\widetilde{D}(u,\tau').
\end{equation}
Finally, taking the limit $x\rightarrow0$ in the double integral, one ends up with the following simple asymptotic behaviour:
\begin{equation}
\bar{k}_\perp^2(x)\simeq \frac{\sqrt{\qhat(L)xE}}{\abar}\int_0^1\dif u\,u\int_0^{\infty}\dif \tau'\,\widetilde{D}(u,\tau')=\frac{\sqrt{\qhat(L)xE}}{4\abar}
\end{equation}
which is shown on Fig.~\ref{Fig:frag-func-2} (dotted curves).

This simple relation enables us to quantify the scaling violation due
to transverse broadening between an expanding medium and the ``equivalent'' static one, in the multiple branching regime:
\begin{equation}
 \frac{\bar{k}_\perp^2(x)}{\bar{k}^{2,\rm
     stat}_\perp(x)}=\sqrt{\frac{\qhat(L)}{\qhat_{\rm
       stat}}}=\frac{2-\gamma}{2}\,\frac{1-\left({t_0}/{L}\right)^{1-\frac{\gamma}{2}}}{1-{t_0}/{L}}\,.
\end{equation}
This ratio is always smaller than $1$ for all $t_0\in[0,L]$ and
$\gamma\in[0,2)$.
\bibliographystyle{utcaps}
\bibliography{refs}

\end{document}